\documentclass[letterpaper,twocolumn,10pt]{article}
\usepackage{usenix2019_v3}

\usepackage[english]{babel}
\usepackage{blindtext}
\usepackage{anyfontsize}
\usepackage{enumitem}
\usepackage{tabularx}
\usepackage{colortbl}
\usepackage{subcaption}
\usepackage{appendix}
\usepackage{pdflscape}
\usepackage{comment}
\usepackage{rotating}
\usepackage{float}
\usepackage{bm}

\usepackage{graphicx}

\usepackage{amsmath}
\usepackage{booktabs}
\usepackage{lastpage}

\usepackage{background}
\backgroundsetup{
  position=current page.east,
  angle=-90,
  nodeanchor=center,
  vshift=-0.6cm,
  hshift=0cm,
  opacity=1,
  scale=1.9,
  contents={Under anonymous submission (Sept. 19, 2019)}
}

\usepackage[labelfont=bf]{caption}

\newcommand\name
{Fugu\xspace}
\newcommand\website{Puffer\xspace}
\usepackage{microtype}
\begin{document}

\date{}

\title{\Large \bf Learning \emph{in situ}: a randomized experiment in video streaming}


\author{
\rm
\setlength{\tabcolsep}{12pt}
\def\arraystretch{1.05}
\begin{tabular}{cccc}
  Francis Y. Yan & Hudson Ayers & Chenzhi Zhu\textsuperscript{$\dagger$} & Sadjad Fouladi \\
  James Hong & Keyi Zhang & Philip Levis & Keith Winstein \\
\end{tabular}\\[3.5ex]
  {\it Stanford University, \textsuperscript{$\dagger$}Tsinghua University}
}

\maketitle

\begin{abstract}
We describe the results of a randomized controlled trial of
video-streaming algorithms for bitrate selection and network
prediction. Over the last eight months, we have streamed 14.2~years of
video to 56,000 users across the Internet. Sessions are randomized in
blinded fashion among algorithms, and client telemetry is recorded for
analysis.

We found that in this real-world setting, it is difficult for
sophisticated or machine-learned control schemes to outperform a
``simple'' scheme (buffer-based control), notwithstanding
good performance in network emulators or simulators. We performed a statistical analysis
and found that the variability and heavy-tailed nature of
network and algorithm behavior create hurdles for robust learned
algorithms in this area.

We developed an ABR algorithm that robustly outperforms other
schemes in practice, by combining classical control with a learned network
predictor, trained with supervised learning \emph{in situ} on data
from the real deployment environment.

To support further investigation, we are publishing an archive of
traces and results each day, and will open our ongoing study to the
community. We welcome other researchers to use this platform to
develop and validate new algorithms for bitrate selection, network
prediction, and congestion control.

\end{abstract}

\maketitle

\section{Introduction}

Video streaming is the predominant Internet application, making up
almost three quarters of all traffic~\cite{cisco2017}.  One key
algorithmic question in video streaming is \emph{adaptive bitrate
  selection}, or ABR, which decides the compression level selected for
each ``chunk,'' or segment, of the video. ABR algorithms optimize the
user's quality of experience (QoE): more-compressed chunks reduce
quality, but larger chunks may stall playback if the client cannot
download them in time.


In the academic literature, many recent ABR algorithms use statistical
and machine-learning methods~\cite{mpc,bola,sabre,cs2p,oboe,pensieve},
which allow algorithms to consider many input signals and try to
perform well for a wide variety of clients. An ABR decision can depend
on recent throughput, client-side buffer occupancy, delay, the
experience of clients on similar ISPs or types of connectivity,
etc. Machine learning can find patterns in seas of data and is a
natural fit for this problem domain.

However, it is a perennial lesson that the performance of learned
algorithms depends greatly on the data or environments used to train
them. Internet services revolutionized machine learning in part
because they have huge, rich datasets. Generally speaking, the
machine-learning community has gravitated towards simpler algorithms
trained with huge amounts of representative data---these algorithms
tend to be more robust in novel
scenarios~\cite{ml-data-more-than-algo,revisiting-ml-data}---and away
from sophisticated algorithms trained on small datasets.

Unfortunately, ML approaches for video streaming and other networking
areas are often hampered in their access to good and representative
training data. The Internet is complex and diverse, individual nodes
only observe a noisy sliver of the system dynamics, and behavior is
often heavy-tailed. Accurately simulating or emulating the diversity of
Internet paths remains beyond current
capabilities~\cite{cantsimulateinternet, simulationdifficulties,
  bettermodels, pantheon}.

As a result, algorithms trained in emulated environments may not
generalize to the Internet~\cite{tracebased}. For example, CS2P's gains were more modest
over real networks than in simulation~\cite{cs2p}. Measurements of Pensieve~\cite{pensieve} saw narrower benefits
on similar network paths~\cite{pensieve-reproduce} or large-scale
streaming services~\cite{realworldpensieve}. Other learned algorithms,
such as the Remy congestion-control schemes,
have also seen inconsistent results on real networks, despite good
results in simulation~\cite{pantheon}.

This paper seeks to answer: \emph{what does it take to create a
  learned ABR algorithm that robustly performs well over the wild
  Internet?} We report the design and findings of \website
  \footnote{\url{https://puffer.stanford.edu}}, an
ongoing research study that operates a video-streaming website open to
the public. Over the past eight months, \website has streamed
14.2~years of video to 56,000 distinct users, while recording client
telemetry for analysis (current load is about 50 stream-days of data
per day). \website randomly assigns each session to one of a set of
ABR algorithms; users are blinded to the assignment. We find:

\vspace{1ex}\noindent {\bf In our real-world setting, sophisticated
  algorithms based on control theory~\cite{mpc} or reinforcement
  learning~\cite{pensieve} did not outperform simple buffer-based
  control~\cite{bufferbased}.} We found that more-sophisticated
algorithms do not necessarily beat simpler, older algorithms. The
newer algorithms were developed using collections of
traces that may not have captured enough of the variability
or heavy tails we see in practice.

\vspace{1ex}\noindent {\bf Statistical margins of error in quantifying
  algorithm performance are considerable.}  Prior work on ABR
algorithms has claimed benefits of 10--15\%~\cite{mpc},
3.2--14\%~\cite{cs2p}, or 12--25\%~\cite{pensieve}, based on traces or
real-world experiments lasting hours or days. However, we found that
the empirical variability and heavy tails of throughput evolution and
rebuffering create statistical margins of uncertainty that make it
challenging to detect real effects of this magnitude. Even with a
\emph{year} of accumulated experience (or representative
traces) per scheme, a 20\% improvement in rebuffering ratio would be
statistically indistinguishable, i.e., below the threshold of
detection with 95\% confidence. These uncertainties affect the design
space of machine-learning approaches that can practically be deployed
in this setting~\cite{rlchallenges, reducingvariance}.

\vspace{1ex}\noindent {\bf It is possible to robustly outperform
  existing schemes by combining classical control with an ML predictor
  trained \emph{in situ} on real data.}  We describe \name, a
data-driven ABR algorithm that combines several techniques. \name is
based on MPC (model-predictive control)~\cite{mpc}, a classical
control policy, but replaces its throughput predictor with a deep
neural network trained using supervised learning on data recorded ``in
situ,'' meaning from \name's actual deployment environment, \website.
The predictor has some uncommon features: it predicts
\emph{transmission time} given a proposed chunk's filesize
(vs.~estimating throughput), it outputs a probability distribution
(vs.~a point estimate), and it considers low-level congestion-control
statistics among its input signals. Each of these techniques has been explored before,
but \name combines them in a new way. Ablation studies (Section~\ref{s:ttp}) find
each of these techniques to be necessary to \name's performance.

In a rigorous controlled experiment during most of 2019, \name
outperformed existing techniques---including the simple algorithm---in
stall ratio (with one exception), video quality, and the variability
of video quality (Fig.~\ref{fig:summary}). The improvements were
significant both statistically and, perhaps, practically: users who
were randomly assigned to \name (in blinded fashion) chose to continue
streaming for 10--20\% longer, on average, than users assigned to
the other ABR algorithms\footnote{This effect was statistically
  significant but driven solely by users streaming more than 2.5~hours
  of video; we do not fully understand it.}.

\begin{figure}
\begin{centering}
{
  \setlength{\tabcolsep}{2pt}
  \footnotesize
        \begin{tabular}{lcccc}
          \multicolumn{5}{l}{\textbf{\scalebox{0.95}[1.0]{Results of primary experiment (Jan.~19--Aug.~7 \& Aug.~30--Sept.~12, 2019)}}} \\
          \toprule
          Algorithm & Time stalled & Mean SSIM & SSIM variation & Mean duration \\[-1ex]
                    & \tiny \textcolor{darkgray}{(lower is better)} & \tiny \textcolor{darkgray}{(higher is better)} & \tiny \textcolor{darkgray}{(lower is better)} & \tiny \textcolor{darkgray}{(time on site)} \\ \midrule
          \rowcolor{blue!10}\cellcolor{white}\name        & \cellcolor{white}0.12\% & 16.9~dB & 0.68~dB & 32.6~min\\
          MPC-HM~\cite{mpc}                          & 0.25\% & 16.8~dB & 0.72~dB & 27.9~min\\
          BBA~\cite{bufferbased}                             & 0.19\% & 16.8~dB & 1.03~dB & 29.6~min \\
          Pensieve~\cite{pensieve}                        & 0.17\% & 16.5~dB & 0.97~dB & 28.5~min \\
          \scalebox{.9}[0.95]{RobustMPC-HM}                    & \cellcolor{blue!10}0.10\% & 16.2~dB & 0.90~dB & 27.4~min \\
          \bottomrule
      \end{tabular}
}
\caption{\rm In a seven-month randomized controlled trial with blinded
  assignment, the \name scheme outperformed other ABR algorithms. The
  primary analysis includes 458,801 streams played by 44,907 client IP
  addresses (8.5 client-years in total). Uncertainties are shown in
  Figures~\ref{fig:overall} and \ref{fig:watchtime}.}
\label{fig:summary}
\end{centering}
\end{figure}

\vspace{1ex}

Our results suggest that, as in other domains, good and representative training is
the key distinguishing feature required for robust performance of
learned ABR algorithms. The simplest way to obtain representative training
data is to learn \emph{in situ}, on real data from the actual
deployment environment, assuming the scheme can be trained on observed
data and the deployment is widely enough used to exercise a broad
range of scenarios. The approach we describe here is only a step in
this direction, but we believe \website's results suggest that
machine-learned networking systems will benefit by addressing the
challenge of ``\emph{how will we get enough representative scenarios
  for training---what is enough, and how do we keep them
  representative over time?}''  as a first-class consideration.

We intend to operate \website as an ``open research'' project for the next several years. We invite the
research community to train and test new algorithms on randomized
subsets of its traffic, gaining feedback on real-world
performance with quantified uncertainty. Along with this paper, we are
publishing an archive of traces and results back to the beginning of
2019, with new traces and results posted daily.

In the next few sections, we discuss the background and related work
on this problem (\S\ref{s:relwork}), the design of our blinded
randomized experiment (\S\ref{s:ourtube}) and the \name
algorithm (\S\ref{s:design}), with experimental results in
Section~\ref{s:evaluation}, and a discussion of results and
limitations in Section~\ref{s:limits}. In the appendices, we provide a
standardized diagram of the experimental flow for the primary analysis
and describe the format and quantity of data we are releasing
alongside this paper.

\section{Background and Related Work}
\label{s:relwork}

The basic problem of adaptive video streaming has been the subject of
much academic work; for a good overview, we refer the reader to Yin et
al.~\cite{mpc}. We briefly outline the problem here. A service wishes
to serve a pre-recorded or live video stream to a broad array of
clients over the Internet. Each client's connection has a different
and unpredictable time-varying performance. Because there are many
clients, it is not feasible for the service to adjust the encoder
configuration in real time to accommodate any one client.

Instead, the service encodes the video into a handful of alternative
compressed versions. Each represents the original video but at a
different quality, target bitrate, or resolution. Client sessions
choose from this limited menu. The service encodes the different
versions in a way that allows clients to switch midstream as
necessary: it divides the video into \emph{chunks}, typically 2--6
seconds each, and encodes each version of each chunk independently, so
it can be decoded without access to any other chunks. This gives
clients the opportunity to switch between different versions at each
chunk boundary. The different alternatives are generally referred to
as different ``bitrates,'' although streaming services today generally
use ``variable bitrate'' (VBR) encoding~\cite{qin2018abr}, where
within each alternative stream, the chunks vary in compressed
size~\cite{zhang2017modeling}.

\vspace{0.5\baselineskip}
\noindent \textbf{Choosing which chunks to fetch.} Algorithms that
select which alternative version of each chunk to fetch and play,
given uncertain future throughput, are known as {\it adaptive bitrate}
(ABR) schemes. These schemes fetch chunks, accumulating them in a
playback buffer, while playing the video at the same time. The
playhead advances and drains the buffer at a steady rate, 1 s/s, but
chunks arrive at irregular intervals dictated by the varying network
throughput and the compressed size of each chunk. If the buffer
underflows, playback must stall while the client
``rebuffers'': fetching more chunks before resuming playback.  The
goal of an ABR algorithm is typically framed as choosing the optimal
sequence of chunks to fetch or replace~\cite{sabre}, given recent experience and guesses about the future, to
minimize startup time and presence of stalls,
maximize the quality of chunks played back, and minimize
variation in quality over time (especially abrupt changes in
quality). The importance tradeoff for these factors is captured in a
QoE metric; several studies have
calibrated QoE metrics against human behavior or
opinion~\cite{qoe-waterloo, ramesh2012,predictivemodel}.

\vspace{0.5\baselineskip}
\noindent \textbf{Adaptive bitrate selection.}  Researchers have
produced a literature of ABR schemes, including ``rate-based''
approaches that focus on matching the video bitrate to the network
throughput~\cite{festive, probeandadapt, qdash}, ``buffer-based''
algorithms that steer the duration of the playback
buffer~\cite{bufferbased, bola, sabre}, and control-theoretic schemes
that try to maximize expected QoE over a receding horizon, given the
upcoming chunk sizes and a prediction of the future throughput.

Model Predictive Control (MPC), FastMPC, and Robust\-MPC~\cite{mpc}
fall into the last category. They comprise two modules: a \emph{throughput
  predictor} that
informs a predictive \emph{model} of what will happen to the buffer occupancy and
QoE in the near future, depending on which chunks it fetches, with what quality
and sizes. MPC uses the model to plan a sequence of
chunks over a limited horizon (e.g.,~the next 5--8 chunks) to
maximize the expected QoE. We implemented MPC and RobustMPC for
\website, using the same predictor as the paper: the harmonic mean of
the last five throughput samples.

\begin{figure}
\begin{subfigure}{0.48\columnwidth}
\includegraphics[width=\textwidth]{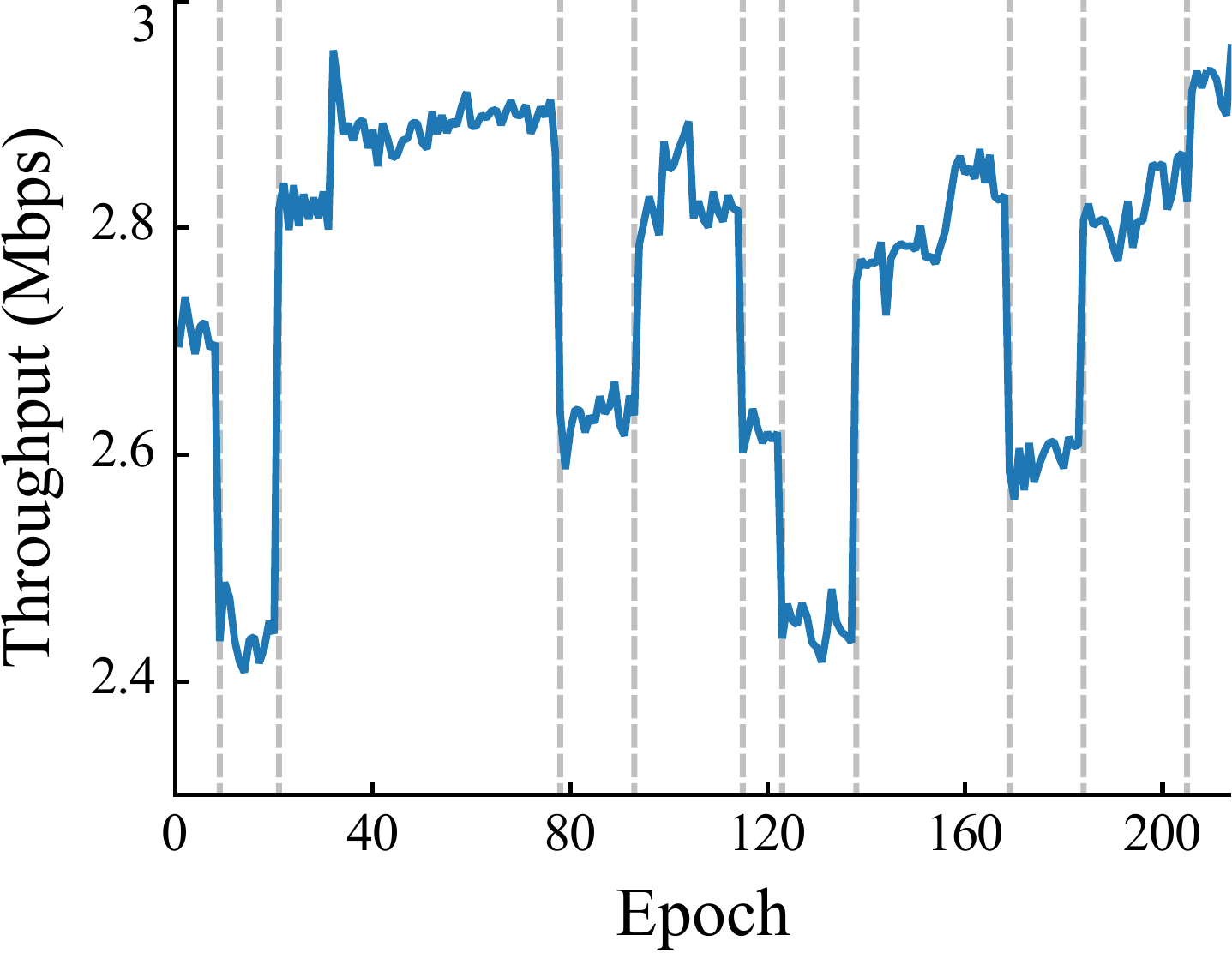}
\caption{\rm CS2P example session (Figure~4a from~\cite{cs2p})}
\end{subfigure}
\hfill
\begin{subfigure}{0.48\columnwidth}
\includegraphics[width=\textwidth]{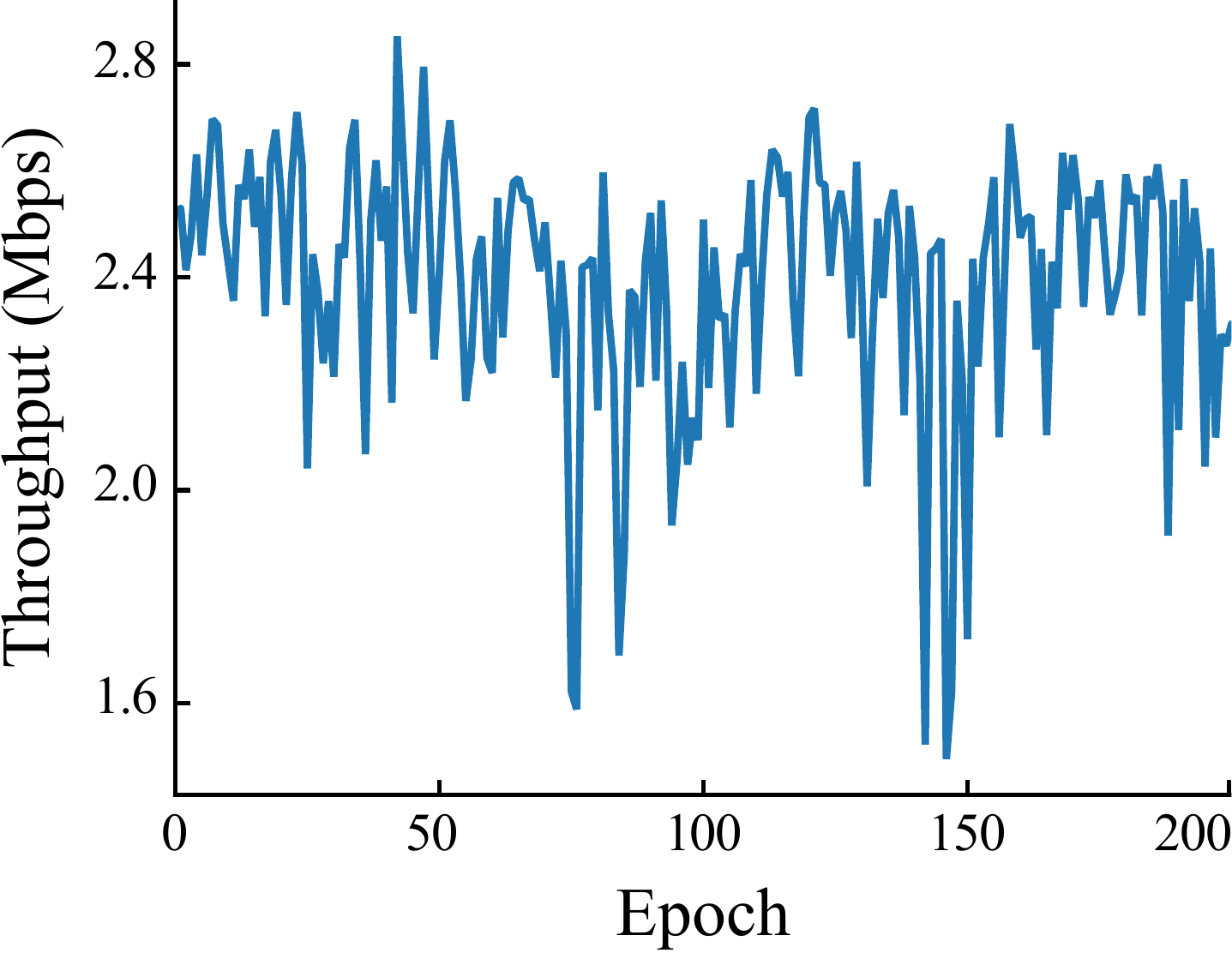}
\caption{\rm Typical \website session with similar mean throughput}
\end{subfigure}
\caption{\rm \website has not observed CS2P's discrete throughput states. (Epochs are 6 seconds in both plots.)}
\label{fig:markovian}
\end{figure}

CS2P~\cite{cs2p} and Oboe-tuned RobustMPC~\cite{oboe} are related
to MPC; they constitute better throughput predictors that inform the
same control strategy (MPC).  These throughput predictors were trained
on real datasets that recorded the evolution of throughput over time
within a session; CS2P clusters users by similarity and models their
evolving throughput as a Markovian process with a small number of
discrete states; Oboe uses a similar model to detect when the network
path has changed state. In our dataset, we have not observed CS2P and
Oboe's observation of discrete throughput states
(Figure~\ref{fig:markovian}).

\name fits in this same category of algorithms. It also uses MPC as
the control strategy, informed by a network predictor trained on real
data. This component, which we call the Transmission Time Predictor
(TTP), incorporates a number of features, none of which can claim
novelty on its own. The TTP is not a ``throughput'' predictor
\emph{per se}; it predicts the transmission time of a proposed chunk
with a given filesize. That observed throughput varies with filesize
is a well-known effect~\cite{tracebased, qin2018abr,
  zhang2017modeling}, although to our knowledge \name is the first to
use this fact operationally as part of a control policy. \name's
predictor is also \emph{probabilistic}---it outputs not a single
predicted transmission time, but a probability distribution on
possible outcomes. The use of probabilistic or stochastic uncertainty
in model predictive control has a long history~\cite{chancempc}, but
to our knowledge \name is the first to use stochastic MPC in this
context. Finally, \name's predictor is a neural network, which lets it
consider an array of diverse signals that relate to transmission
time, including raw congestion-control statistics from the sender-side
TCP implementation~\cite{tian2012towards, salsify}. We found that
several of these signals (RTT, delivery\_rate, FlightSize) benefit ABR
decisions (\S\ref{s:evaluation}).

Pensieve~\cite{pensieve} is an ABR scheme also based on a deep neural
network. Unlike \name, Pensieve uses the neural network not simply to
make predictions but to make \emph{decisions} about which chunks to
send.  This affects the type of learning used to train the
algorithm. While CS2P and \name's TTP can be trained with
\emph{supervised learning} (to predict chunk transmission times
recorded from past data), it takes more than ``data'' to train a
scheme that makes decisions; these schemes need training
\emph{environments} that respond to a series of decisions and judge
their consequences. This is known as ``reinforcement learning.''
Generally speaking, reinforcement learning techniques need to be able
to observe a detectable difference in performance by slightly varying
a control action; this requires large amounts of training, and systems
that are challenging to simulate faithfully or that have too much
variability present difficulties~\cite{rlchallenges,
  reducingvariance}. The authors of Pensieve recently tested a similar
scheme on video traffic at Facebook~\cite{realworldpensieve},
observing a 1.6\% increase in bitrate in a large real-world test.

\section{\website: an ongoing live study of ABR}
\label{s:ourtube}

To understand the challenges of video streaming and measure the behavior
of ABR schemes, we built \website, a free, publicly accessible website
that live-streams six over-the-air commercial television
channels. \website operates as a randomized controlled trial; sessions
are randomly assigned to one of a set of ABR or congestion-control
schemes. The study participants include any member of the public who
wishes to participate. Users are blinded to algorithm assignment,
and we record client telemetry on video quality and playback.  Our
Institutional Review Board determined that \website does not
constitute human subjects research.

Our reasoning for streaming live television was to collect data from enough
participants and network paths to draw robust conclusions about the
performance of algorithms for ABR control and network prediction. Live
television is an evergreen source of popular content
that had not been broadly available for free on the
Internet. Our study benefits, in part, from a law that allows
nonprofit organizations to retransmit over-the-air television signals
without charge~\cite{locastwhitepaper}. Here, we describe details
of the system, experiment, and analysis.

\begin{figure}
\begin{subfigure}[t]{0.48\columnwidth}
\includegraphics[height=0.72\textwidth]{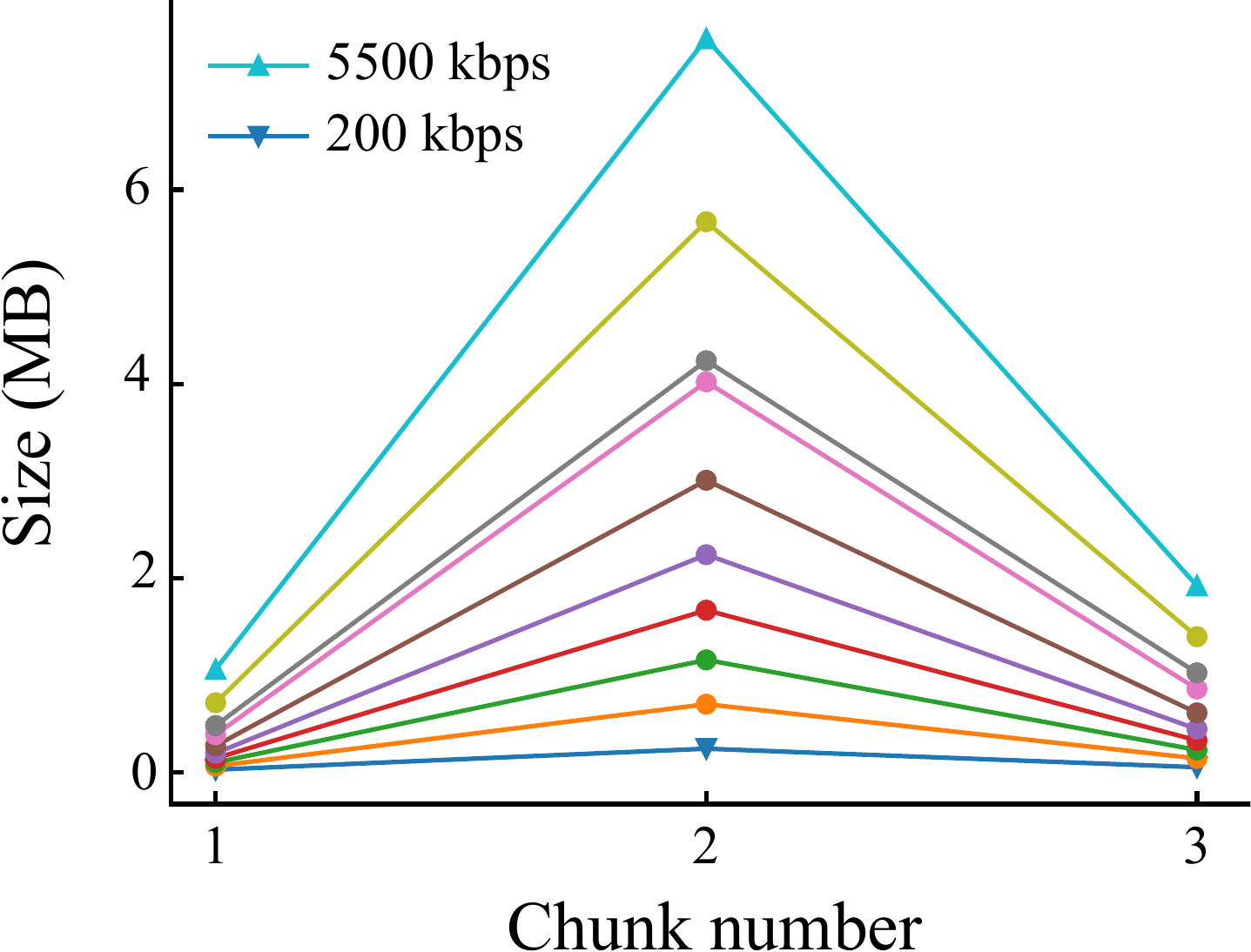}
\caption{\rm VBR encoding lets chunk size vary within a stream~\cite{zhang2017modeling}.}
\end{subfigure}%
\hfill
\begin{subfigure}[t]{0.48\columnwidth}
\includegraphics[height=0.72\textwidth]{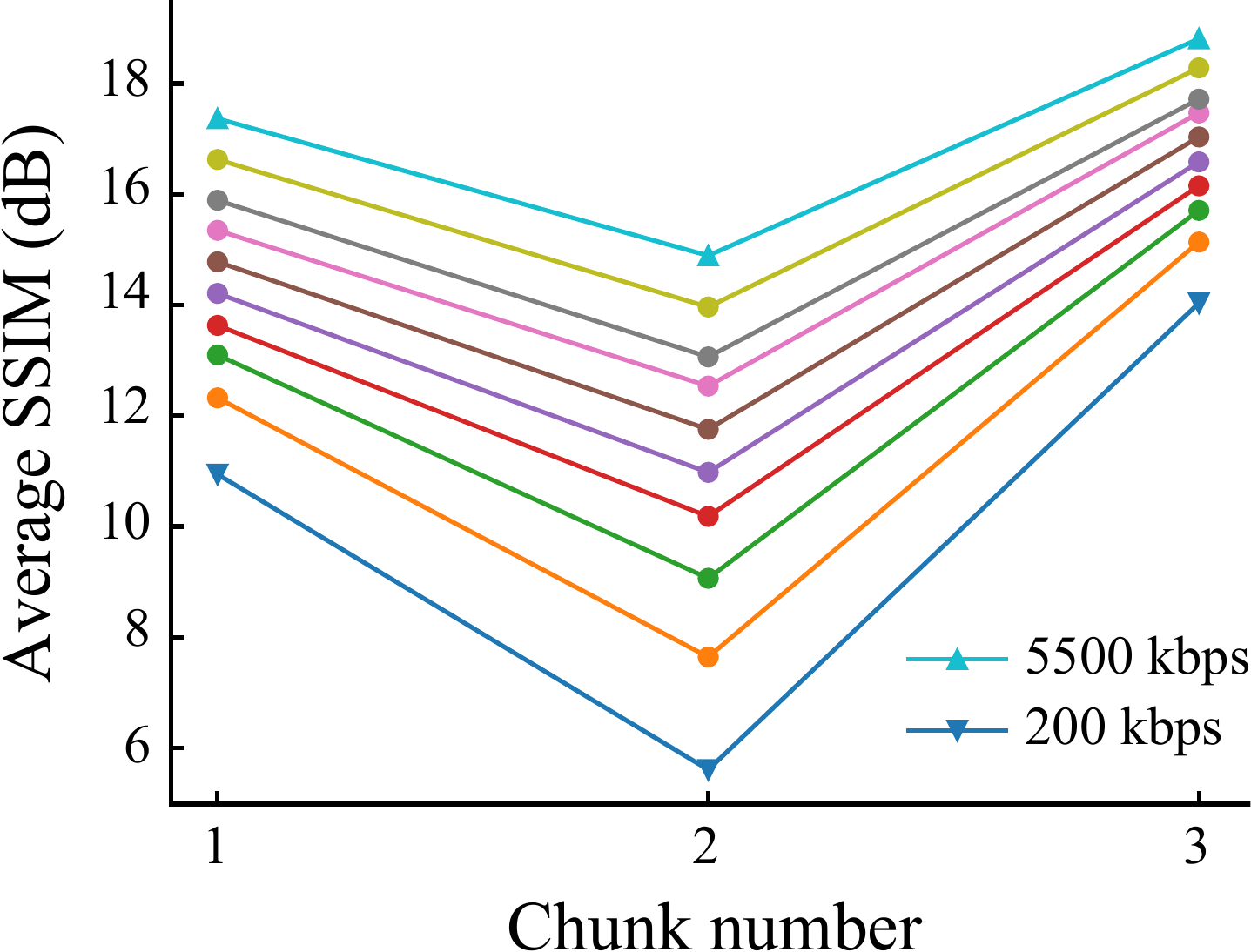}
\caption{\rm Picture quality also varies with VBR encoding~\cite{qin2018abr}.}
\end{subfigure}
\caption{\rm Variations in picture quality and chunk size within each stream
  suggest a benefit from choosing chunks based on SSIM and size, rather than average
  bitrate (legend).}
\label{fig:ssim_v_size}
\end{figure}

\subsection{Back-end: decoding, encoding, SSIM}

\website receives six television channels using a VHF/UHF antenna and
an ATSC demodulator, which outputs MPEG-2 transport streams in UDP. We
wrote software to decode a stream to chunks of raw decoded video and
audio, maintaining synchronization (by inserting black fields or
silence) in the event of lost transport-stream packets on either
substream. Video chunks are 2.002 seconds long, reflecting the 1/1001
factor for NTSC frame rates. Audio chunks are 4.8 seconds long.  Video
is de-interlaced with \texttt{ffmpeg} to produce a ``canonical''
1080p60 or 720p60 source for compression.

\website encodes each video chunk in ten different H.264
versions, using \texttt{libx264} in \texttt{veryfast} mode. The encodings range from 240p60
video with constant rate factor (CRF) of 26 (about 200~kbps) to
1080p60 video with CRF of 20 (about 5,500~kbps). Audio chunks are
encoded in the Opus format.

\begin{figure}
  \begin{centering}
    \includegraphics[width=0.85\columnwidth]{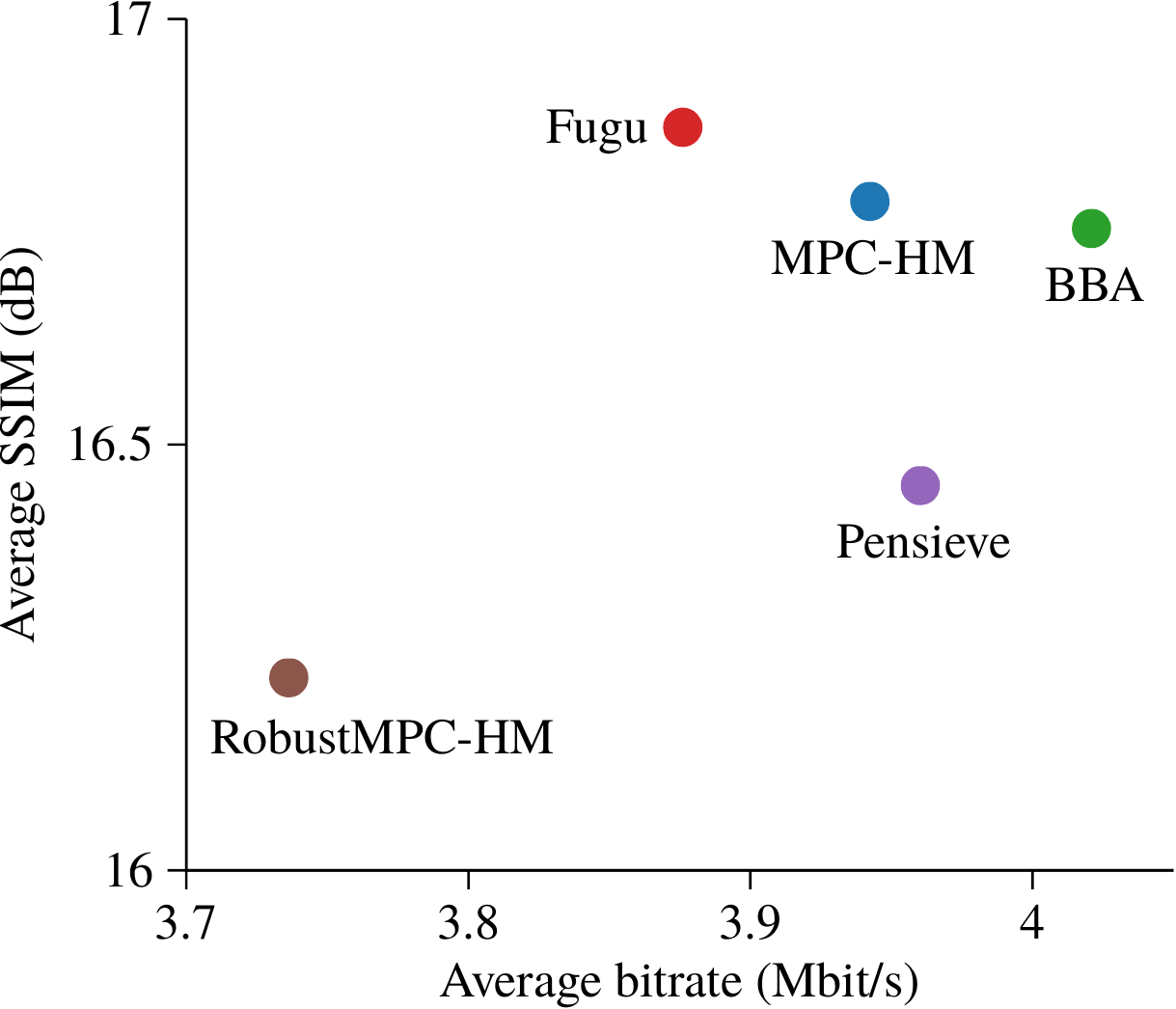}

    \end{centering}
  \caption{\rm On \website, schemes that maximize average SSIM (MPC-HM, RobustMPC-HM, and \name)
    delivered higher quality video per byte sent, vs.~those that maximize bitrate directly (Pensieve)
    or the SSIM of each chunk (BBA).}
\label{fig:bitrate}
\end{figure}

\website then uses \texttt{ffmpeg} to calculate each encoded chunk's
SSIM~\cite{ssim}, a measure of video quality, relative to the
canonical source. This information is used by the objective function
of BBA, MPC, RobustMPC, and \name, and for our evaluation. In
practice, the relationship between bitrate and quality varies
chunk-by-chunk (Figure~\ref{fig:ssim_v_size}), and users cannot
perceive compressed chunk sizes directly---only what is shown on the
screen. Our results indicate that schemes that maximize bitrate
directly do not reap a commensurate benefit in picture quality
(Figure~\ref{fig:bitrate}).

Encoding six channels in ten versions each (60 streams total) with
\texttt{libx264} consumes about 48 cores of an Intel x86-64 2.7 GHz
CPU in steady state. Calculating the SSIM of each encoded chunk
consumes an additional 18 cores.

\subsection{Serving chunks to the browser}

\newcommand{\boldplus}{\textbf{+}\nobreak\hspace{0.1em}}
\newcommand{\boldminus}{\textbf{--}\nobreak\hspace{0.1em}}

\begin{figure*}
\begin{centering}
\scalebox{0.92}{
\begin{tabular}{@{}lllll@{}}
\toprule
Algorithm & Control & Predictor & Optimization goal & How trained \\
\midrule
BBA & classical (prop.~control) & \textit{\textcolor{black!50!white}{n/a}} & \boldplus SSIM \emph{s.t.}~bitrate \textbf{<} limit & \textit{\textcolor{black!50!white}{n/a}} \\
MPC-HM    & classical (MPC) & classical (HM) & \boldplus $\overline{\textrm{SSIM}}$, \boldminus stalls, \boldminus$\Delta$SSIM & \textit{\textcolor{black!50!white}{n/a}} \\
RobustMPC-HM & classical (robust MPC) & classical (HM) & \boldplus $\overline{\textrm{SSIM}}$, \boldminus stalls, \boldminus$\Delta$SSIM & \textit{\textcolor{black!50!white}{n/a}} \\
Pensieve & learned (DNN) & \textit{\textcolor{black!50!white}{n/a}} & \boldplus \scalebox{0.91}[1.0]{bitrate}, \boldminus stalls, \boldminus$\Delta$bitrate & {\raggedright reinforcement learning in simulation} \\
\scalebox{0.94}[1.0]{Emulation-trained \name{}}  & classical (MPC) & learned (DNN) & \boldplus $\overline{\textrm{SSIM}}$, \boldminus stalls, \boldminus$\Delta$SSIM & {\raggedright supervised learning in emulation} \\
\name{}  & classical (MPC) & learned (DNN) & \boldplus $\overline{\textrm{SSIM}}$, \boldminus stalls, \boldminus$\Delta$SSIM & {\raggedright supervised learning \emph{in situ}} \\
\bottomrule
\end{tabular}}
\end{centering}

\caption{\rm Distinguishing features of algorithms used in the experiments. HM = harmonic mean of last five
throughput samples. MPC = model-predictive control. DNN = deep neural network.}
\end{figure*}

To make it feasible to deploy and test arbitrary ABR schemes, \website
uses a ``dumb'' player (using the HTML5 \texttt{<video>} tag and the
JavaScript MediaSource extensions) on the client side, and places the
ABR scheme at the server. We have a 48-core server with 10~Gbps
Ethernet in a well-connected datacenter. The browser opens a WebSocket
(TLS/TCP) connection to a daemon on the server. Each daemon is
configured with a different TCP congestion control (for the primary
analysis, we used BBR~\cite{bbr}) and ABR scheme. Some schemes are
more efficiently implemented than others; on average the CPU load from
serving client traffic (including TLS, TCP, and ABR) is about 5\% of
an Intel x86-64 2.7~GHz core per stream.  Sessions are randomly
assigned to serving daemons. Users can switch channels without
breaking their TCP connection and may have many ``streams'' within
each session.


\website is not a client-side DASH~\cite{mpeg-dash}
(Dynamic Adaptive Streaming over HTTP) system. Like DASH, though, \website
is an ABR system streaming chunked video over a TCP connection, and
runs the same ABR algorithms that DASH systems can run. We don't
expect this architecture to replace client-side ABR (which can be
served by CDN edge nodes), but we expect its conclusions to translate
to ABR schemes broadly. The \website website works in the Chrome,
Firefox, and Edge browsers, including on Android phones, but does not
play in the Safari browser or on iOS (which lack support for
the MediaSource extensions or Opus audio).

\subsection{Hosting arbitrary ABR schemes}

We implemented buffer-based control (BBA), MPC, RobustMPC, and \name
in back-end daemons that serve video chunks over the WebSocket. We use
SSIM in the objective functions for each of
these schemes. For BBA, we used the formula in the original
paper~\cite{bufferbased} to choose reservoir values consistent with
a 15-second maximum buffer.

\vspace{0.5\baselineskip} \textbf{Deploying Pensieve for live
  streaming.}  We use the released Pensieve code (written in Python
with TensorFlow) directly. When a client is assigned to Pensieve,
\website spawns a Python subprocess running Pensieve's multi-video
model.

We contacted the Pensieve authors to request advice on deploying the
algorithm in a live, multi-video, real-world setting.  The authors
recommended that we use a longer-running training and that we tune the
entropy parameter when training the multi-video neural network. We
wrote an automated tool to train 6 different models with various
entropy reduction schemes. We tested these manually over a few real
networks, then selected the model with the best performance.  We
modified the Pensieve code to set {\small \texttt{video\_num\_chunks}}
to 43200, indicating 24 hours of video, so that Pensieve does not
expect the video to end before a user's session completes. We were not
able to modify Pensieve to optimize SSIM or to consider the individual
filesizes of each chunk; it considers the average bitrate of each
\website stream. We adjusted the video chunk length to 2.002 seconds and
the buffer threshold to 15 seconds to reflect our parameters. For
training data, we used the authors' provided script to generate 1000
simulated videos as training videos, and a combination of the FCC and
Norway traces linked to in the Pensieve codebase as training traces.

\subsection{The \website experiment}

To recruit participants, we purchased Google and Reddit ads for
keywords such as ``live tv'' and ``tv streaming,'' and paid people on
Amazon Mechanical Turk to stream video from \website. We were featured
in press articles as a way to watch popular live events (including the
Super Bowl, the World Cup, and other sporting events, ``Bachelor in
Paradise,'' etc.). Our current average load is about 50 stream-days
per day. Popular events bring large spikes ($>20\times$)
over baseline load.

Starting from the beginning of 2019, we have streamed 14.2 years of
video to 55,897 registered study participants using 61,682 unique IP
addresses. About seven months of that period was spent on a randomized
trial comparing \name with other algorithms (MPC, RobustMPC, Pensieve,
and BBA); we refer to this as the primary experiment. This period saw
337,170 streaming sessions, and a total of 1,595,356 individual
streams. A full experimental-flow diagram in the standardized CONSORT
format~\cite{schulz2010consort} is in the appendix
(Figure~\ref{fig:consort}).

\paragraph{Metrics and statistical uncertainty.} We record throughput traces and client telemetry (a full description is in the appendix)
and calculate a set of figures to summarize each stream: the total
time between the first and last recorded events of the stream, the
startup time, the total watch time between the first and last
successfully played portion of the stream, the total time the video is
stalled for rebuffering, the average SSIM, and the chunk-by-chunk
variation in SSIM. The ratio between ``total time stalled'' and
``total watch time'' is known as the rebuffering ratio or stall ratio,
and is widely used to summarize the performance of streaming video
systems~\cite{quic}.

We observe considerable heavy-tailed behavior in most of these
statistics.  Watch times are skewed (Fig.~\ref{fig:watchtime}),
and rebuffering, while important to QoE, is a rare phenomenon. Of the
458,801 eligible streams considered for the primary analysis across
all five ABR schemes, only 15,788 (3\%) of those streams had
\emph{any} stalls, mirroring commercial services~\cite{quic}.

These skewed distributions create more room for the play of chance to
corrupt the bottom-line statistics summarizing a
scheme's performance---even two identical schemes will see considerable variation
in average performance until a substantial amount of data is
assembled. In this study, we worked to quantify the statistical
uncertainty that can be attributed to the play of chance in assigning
sessions to ABR algorithms. We calculate confidence intervals on
rebuffering ratio with the bootstrap method~\cite{bootstrap},
simulating streams drawn empirically from each scheme's observed
distribution of rebuffering ratio as a function of stream duration. We
calculate confidence intervals on average SSIM using the formula for
weighted standard error, weighting each stream by its duration.

These practices result in substantial confidence intervals: with 1.75
years of data for each scheme, the width of the 95\% confidence
interval on a scheme's stall ratio is between $\pm10\%$ and $\pm17\%$
of the mean value. This is comparable to the magnitude of the total
benefit reported by some academic work that used much shorter
real-world experiments. Even a recent study of a Pensieve-like scheme
on Facebook, which collected data on 30 million streams, did not
detect a change in rebuffering ratio outside the level of statistical
noise.

We conclude that considerations of uncertainty in real-world learning
and experimentation, especially given uncontrolled data from the
Internet with real users, deserve further study. Strategies to import
real-world data into repeatable emulators~\cite{pantheon} or reduce
their variance~\cite{reducingvariance} will likely be helpful in producing robust
learned networking algorithms.

\section{\mbox{\name: design and implementation}}
\label{s:design}

\name is a control algorithm for bitrate selection, designed to be
feasibly trained in place (in situ) on a real deployment environment.
It consists of a classical controller (model predictive control, the
same as in MPC-HM), informed by a nonlinear predictor that can
be trained with supervised learning.

\begin{figure}[ht!]
\centering
\includegraphics[width=\columnwidth]{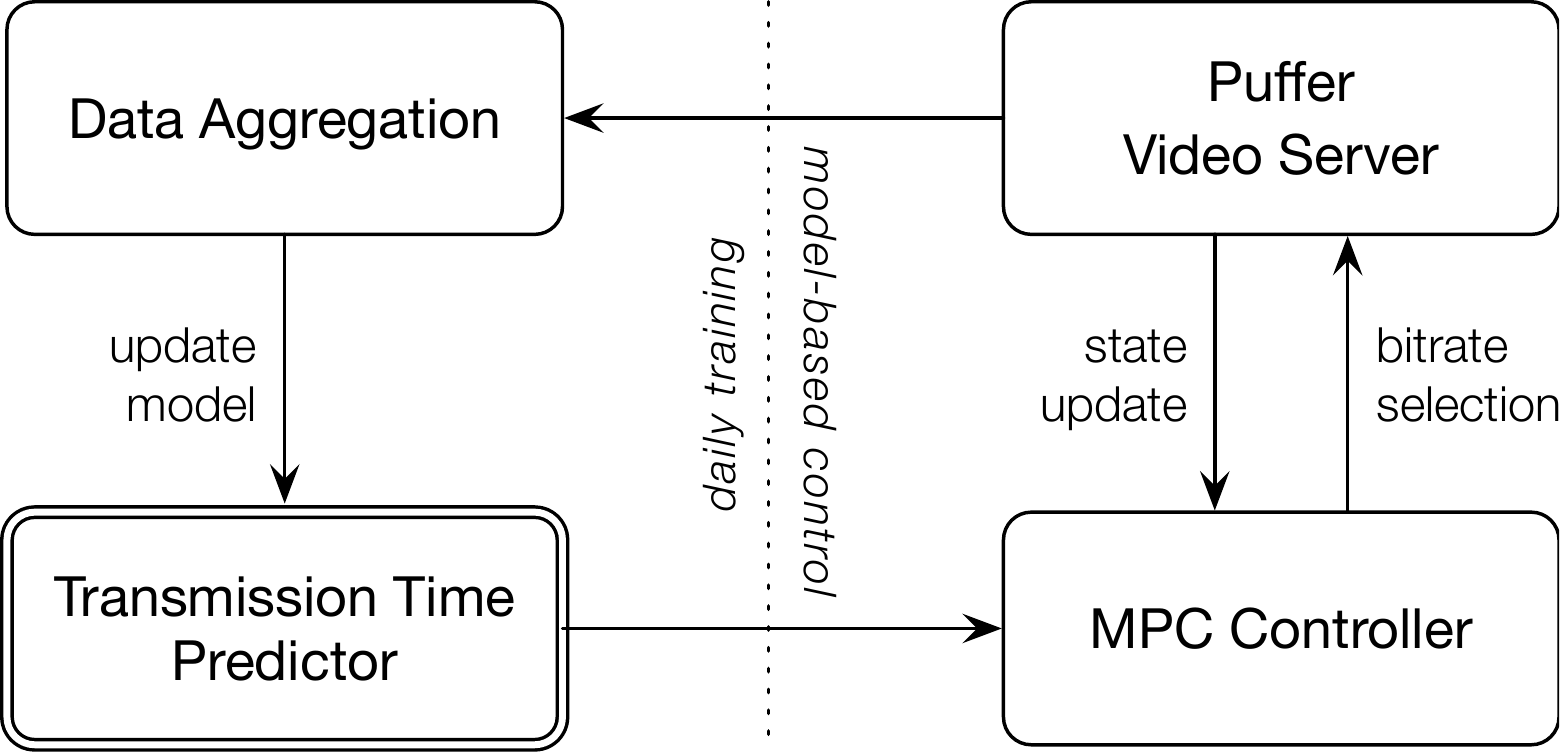}
\caption{\rm Overview of \name}
\label{fig:overview-diagram}
\end{figure}

Figure~\ref{fig:overview-diagram} shows \name's high-level
design. \name runs on the server, making it easy to update its
model and aggregate performance data across clients over time. Clients
send necessary telemetry, such as buffer levels, to the server.


The controller, described in Section~\ref{s:mpc}, makes decisions by
following a classical control algorithm to optimize an objective QoE
function (\S\ref{s:obj}) based on predictions for how long each chunk
would take to transmit. These predictions are provided by the
Transmission Time Predictor (TTP) (\S\ref{s:ttp}), a neural network
that estimates a probability distribution for the transmission
time of a proposed chunk with given filesize.

\subsection{Objective function}
\label{s:obj}

For each video chunk $K_{i}$, \name has a selection of versions
of this chunk to choose from, $K_{i}^{s}$, each with a different size $s$.
As with prior
approaches, \name quantifies the QoE of each chunk as a linear
combination of video quality, video quality variation, and stall
time~\cite{mpc}. Unlike some prior approaches, which use the average
compressed bitrate of each encoding setting as a proxy for image quality, \name optimizes a
perceptual measure of picture quality---in our case, SSIM. This has
been shown to correlate with human opinions of
QoE~\cite{qoe-waterloo}. We emphasize that we use the exact same
objective function in our version of MPC and RobustMPC as well.

Let $Q(K)$ be the quality of a chunk $K$, $T(K)$ be
the uncertain transmission time of $K$, and $B_i$ be the current
playback buffer size. \name defines the QoE of $K_{i}^{s}$ (following \cite{mpc})
as
\begin{equation}\label{eq:qoe}
  \begin{aligned}
    QoE(K_{i}^{s}, K_{i-1}) ={} & Q(K_{i}^{s}) - \lambda |Q(K_{i}^{s}) - Q(K_{i-1})| \\
                                & - \mu \cdot \max\{T(K_{i}^{s}) - B_i, 0\},
  \end{aligned}
\end{equation}
where $\lambda$ and $\mu$ are configuration constants for how much to
weight video quality variation and rebuffering.
The last term $\max\{ T(K_{i}^{s}) - B_i, 0 \}$ describes the stall time experienced
by sending $K_{i}^{s}$. \name
plans a trajectory of sizes $s$ of the future $H$ chunks to
maximize their expected total QoE.

\subsection{Transmission Time Predictor (TTP)}
\label{s:ttp}

Once \name decides which chunk to send, two portions of the QoE
become known: the video quality
and video quality variation. The remaining uncertainty is the
stall time.  The server knows the current playback buffer size,
so what it needs to know is the transmission time: how long
will it take for the client to receive the chunk?
Given an oracle that reports the transmission time of any chunk, the MPC
controller can compute the optimal plan to maximize QoE.

\name uses a trained neural-network transmission-time predictor to
approximate the oracle. For each chunk in the fixed horizon, we train
a separate predictor. E.g., if optimizing for the total QoE of the
next five chunks, five neural networks are trained.  (Multiple
networks in parallel are functionally equivalent to one that
takes the future time step as a variable. We have observed
equivalent performance with both approaches, but training multiple
DNNs lets us parallelize training.)

Each TTP network takes as input a vector of:
\begin{enumerate}[itemsep=0pt]
  \item sizes of past $t$ chunks: $K_{i-t}, \ldots, K_{i-1}$,
  \item transmission times of past $t$ chunks $T_{i-t}, \ldots, T_{i-1}$,
  \item internal TCP statistics (Linux {\small \texttt{tcp\_info}} structure),
  \item size of the chunk to be transmitted.
\end{enumerate}

\noindent The TCP statistics include the current congestion
window size, the number of unacknowledged packets in flight, the
smoothed RTT estimate, the minimum RTT, and the estimated
throughput ({\small \texttt{tcpi\_delivery\_rate}}).

Prior approaches have used Harmonic Mean (HM)~\cite{mpc} or a Hidden
Markov Model (HMM)~\cite{cs2p} to predict a single throughput for the
entire lookahead horizon. In contrast, the transmission-time predictor
outputs a probability distribution $\hat{T}(K_{i}^{s})$ over the
transmission time of $K_{i}^{s}$.

\subsection{Training the TTP}
\label{s:training}

\website collects training data $\mathcal{D}$ by saving client telemetry
from real usage, aggregating pairs of (a) the input 4-vector and, (b) the true
transmission time for the chunk. 
We train the TTP on $\mathcal{D}$ with standard supervised
learning: the training minimizes the cross-entropy loss between the output
probability distribution and the discretized actual transmission time
using stochastic gradient descent.

We retrain the TTP every day, using training data collected on
\website over the prior 14 days, to avoid the effects of dataset shift
and catastrophic forgetting~\cite{ross2011reduction,
  robins1995catastrophic}. Within the 14-day window, we weight more
recent days more heavily, and we shuffle the sampled data to remove
correlation in the sequence of inputs. The weights from the previous
day's model are loaded to warm-start the retraining.


\subsection{Model-based controller}
\label{s:mpc}

Our MPC controller (used for MPC-HM, RobustMPC-HM, and \name) is a stochastic optimal controller that
maximizes the expected cumulative QoE in Equation~\ref{eq:qoe}. It
queries TTP for predictions of transmission time and outputs a plan
$K_{i}^{s}, K_{i+1}^{s}, \ldots, K_{i+H-1}^{s}$ by value
iteration~\cite{bellman1957markovian}. After sending $K_{i}^{s}$, the controller
observes and updates the input vector passed into TTP, and replans
again for the next chunk.

Given the current playback buffer level, let $v^*_i(B_i, K_{i-1})$ denote the
maximum expected sum of QoE that can be achieved in the $H$-step lookahead horizon given
the last sent chunk $K_{i-1}$.
We have value iteration as follows:
\begin{equation*}
  \begin{aligned}
    v^*_i(B_i, K_{i-1}) = \max_{K_{i}^{s}} \Big\{ & \sum_{T_i} \Pr[\hat{T}({K_{i}^{s}}) = T_i] \cdot \\
       & (QoE(K_i^s, K_{i-1}) + v^*_{i+1}(B_{i+1}, K_i^s)) \Big\},
  \end{aligned}
\end{equation*}
where $\Pr[\hat{T}({K_{i}^{s}}) = T_i]$ is the probability for TTP
to output a discretized transmission time $T_i$, and $B_{i+1}$ can
be derived by system dynamics once $T_i$ is estimated.
The controller
computes the optimal trajectory by solving the above value iteration
with dynamic programming (DP). To make the DP computational feasible,
it discretizes $B_i$ into bins and uses forward recursion with memoization to
only compute for relevant $v^*_i(B_i, K_{i-1})$.

\subsection{Implementation}
TTP takes as input the past $t=8$ chunks, and outputs a probability distribution over
21 bins of transmission time: $[0, 0.25), [0.25, 0.75), [0.75, 1.25), \ldots, [9.75, \infty)$,
with 0.5 seconds as the bin size except for the first and the last bins.
TTP is a fully-connected neural network, with two hidden layers
with 64 neurons each. We tested different TTPs with various numbers of hidden layers and
neurons, and found similar training
losses across a range of conditions for each. We implemented TTP and the training
in PyTorch, but we load the trained model in C++ when running on the production server for
performance. A forward pass of TTP's neural network in C++ imposes minimal overhead per chunk
(less than 0.3~ms on average on a recent x86-64 core).
The MPC controller optimizes over $H=5$ future steps (about 10 seconds)
by solving value iteration.  
We set $\lambda=1$ and $\mu=100$ to balance the conflicting goals in QoE. 
Each retraining takes about 6 hours on a 48-core server.

\subsection{Ablation study of TTP features}

\label{ss:ablation}

\begin{figure}
\centering
\includegraphics[width=\columnwidth]{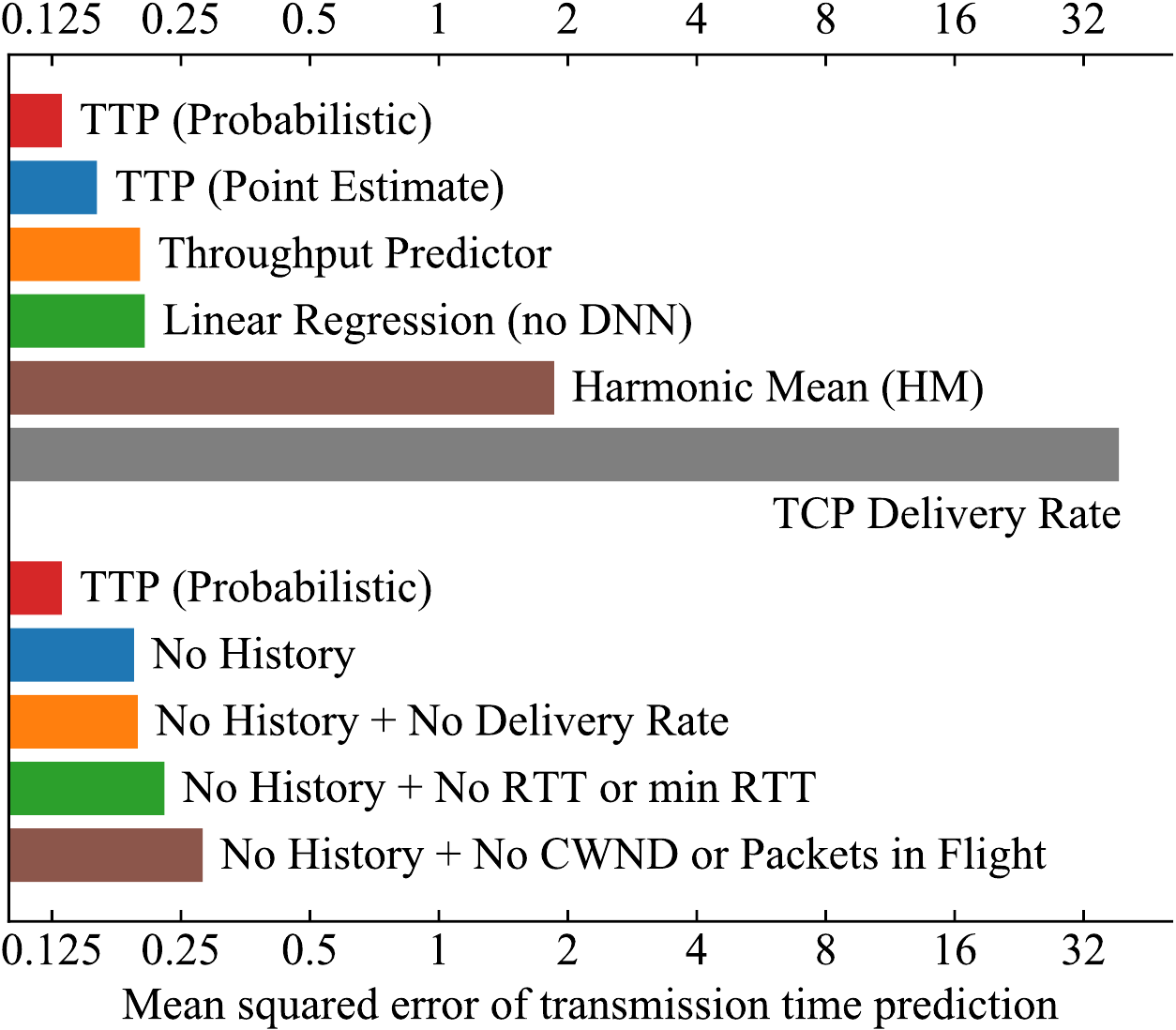}
\caption{\rm Ablation study of \name's Transmission Time Predictor
  (TTP).  Removing each of the TTP's inputs, outputs, or features
  reduced its ability to predict the transmission time of a video
  chunk. A non-probabilistic TTP (``Point Estimate'') and one
  that predicts throughput without regard to chunk size (``Throughput
  Predictor'') both performed markedly worse. TCP-layer
  statistics (RTT, CWND) were also helpful.}
\label{tab:ttp_ablation}
\end{figure}

\begin{figure*}[t]
    \centering
    \begin{subfigure}{1\columnwidth}
      \begin{centering}
        \small\textbf{Primary experiment \textcolor{white!50!black}{(N=458,801, duration 8.5y)}}\\

      \vspace{\baselineskip}

      \includegraphics[width=0.92\columnwidth]{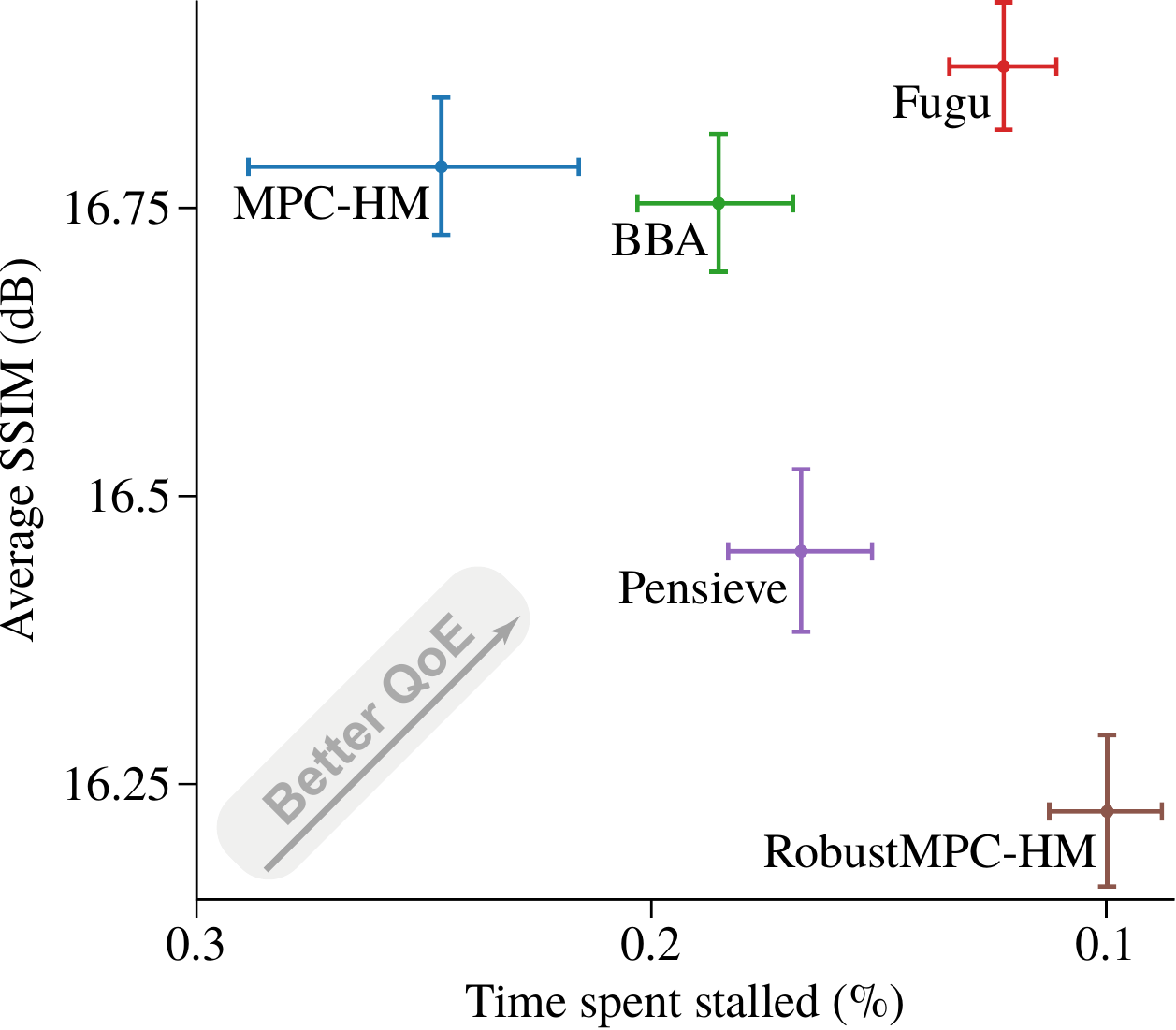}

      \end{centering}
    \end{subfigure}%
    \hfill
    \begin{subfigure}{1\columnwidth}
      \begin{centering}
        \small\textbf{Slow network paths \textcolor{white!50!black}{(N=100,500, duration 1.3y)}}\\

      \vspace{\baselineskip}

      \includegraphics[width=0.92\columnwidth]{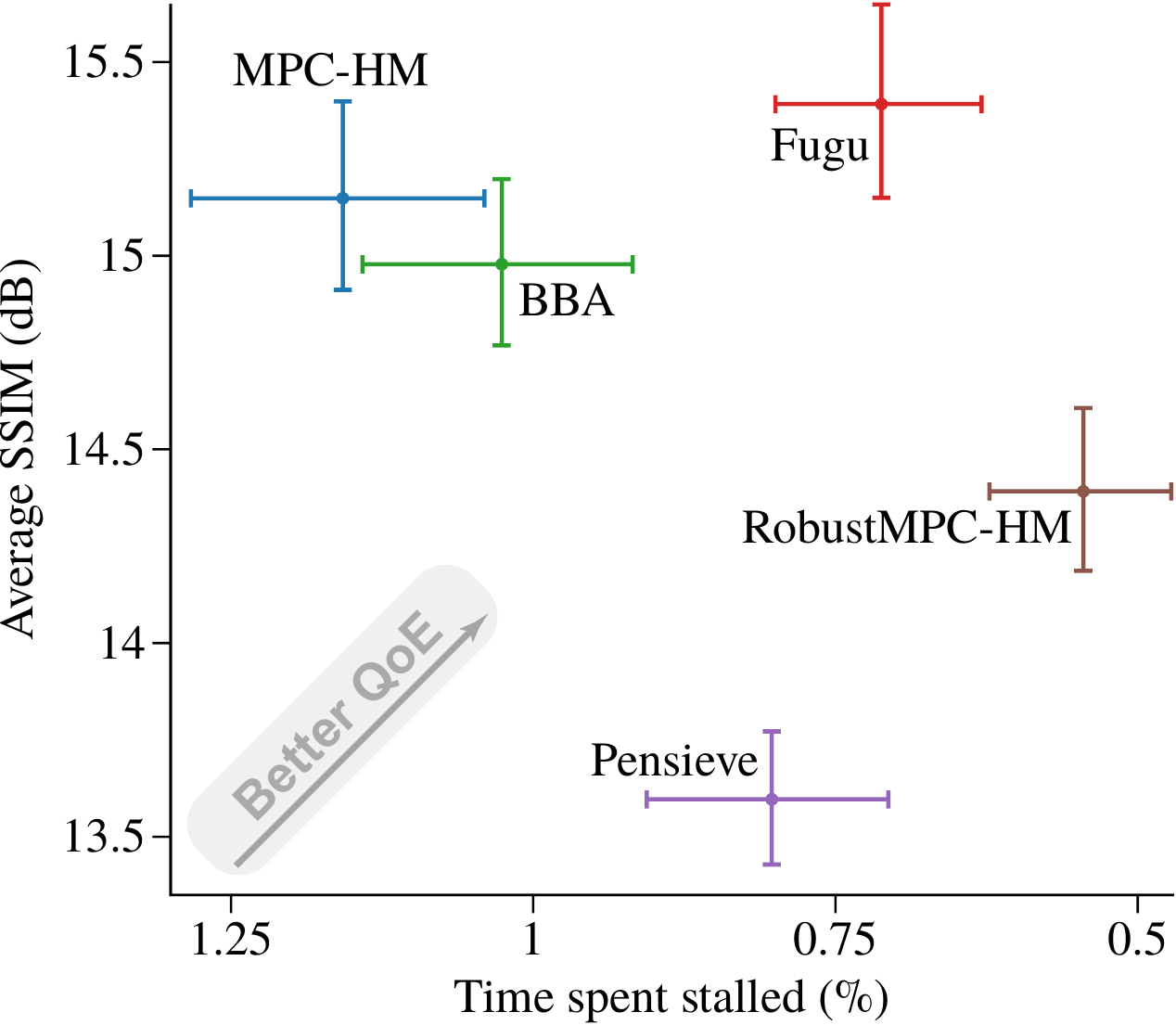}

      \end{centering}
    \end{subfigure}

    \caption{\rm \textbf{Main results.} In a blinded randomized controlled trial that
      included 8.5~years of video streamed to 44,907 client IP
      addresses over a seven-month period, \name reduced the fraction
      of time spent stalled (except with respect to RobustMPC-HM),
      increased SSIM, and reduced SSIM variation within each stream
      (tabular data in Figure~\ref{fig:summary}). ``Slow'' network paths have mean TCP
      \textit{delivery\_rate} less than 6~Mbit/s; following prior
      work~\cite{mpc,pensieve}, these paths are more likely to
      require nontrivial bitrate-adaptation logic. Such streams
      accounted for 16\% of overall viewing time and 82\% of
      stalls. Error bars show 95\% confidence intervals.}
      \label{fig:overall}
      \label{fig:slow_bbr_ssim_rebuf}
\end{figure*}

We performed an ablation study to assess the impact of the TTP's
features (Fig.~\ref{tab:ttp_ablation}). Here are the more
notable results:

\vspace{\baselineskip}
\noindent \textbf{Use of low-level congestion-control statistics.}
The TTP's nature as a DNN lets it consider a variety of noisy inputs, including
low-level congestion-control statistics. We feed the kernel's {\small
  \texttt{tcp\_info}} structure to the TTP,
and find that several of these fields
contribute positively to the TTP's accuracy, especially the RTT, CWND,
and number of packets in flight
(Figure~\ref{tab:ttp_ablation}). Although client-side ABR systems
cannot typically access this structure directory because the
statistics live on the sender, these results should
motivate the communication of richer data to ABR algorithms wherever
they live.

\vspace{\baselineskip}
\noindent \textbf{Transmission-time prediction.} The TTP explicitly considers the chunk size of $K_i$ and outputs a
predicted duration, a more powerful approach than a generating a
single throughput estimate.  It is well known in ABR streaming and
congestion control that transmission time does not scale linearly with
filesize~\cite{tracebased}. We compared the accuracy of the TTP
against an equivalent throughput predictor (keeping everything else
unchanged) and found the TTP's predictions were much more accurate
(Figure~\ref{tab:ttp_ablation}).

\vspace{\baselineskip}
\noindent \textbf{Prediction with uncertainty.} The TTP outputs a \emph{probability distribution} of transmission
times. This additional information allows for better decision making,
compared with a single point estimate without uncertainty. We
evaluated the expected accuracy of a probabilistic TTP vs.~an
equivalent ``maximum likelihood'' version, and found a considerable
improvement in prediction accuracy with the former
(Figure~\ref{tab:ttp_ablation}).

In addition, to confirm the relationship between prediction accuracy
and performance of the entire ABR system, we deployed a point-estimate
version of \name on \website in August 2019 and collected 39
stream-days of data with this scheme. It performed much worse than
normal \name: the rebuffering ratio was 3--9$\times$ worse, without
significant improvement in SSIM (data not shown).

\vspace{\baselineskip}
\noindent \textbf{Use of neural network.} Although it is often said that ``data > algorithms'' in
machine learning~\cite{ml-data-more-than-algo}, we did find a
significant benefit to the use of a deep neural network in this
application. A linear-regression model (equivalent to a single-layer
neural network), trained the same, performs much worse on prediction accuracy
(Figure~\ref{tab:ttp_ablation}).

We also deployed this scheme on the \website website and collected 107
stream-days of data in September 2019 to measure its end-to-end ABR
performance. Again, the lower prediction accuracy was harmful to the
bottom line; its rebuffering ratio was 2--5$\times$ worse (data not
shown).




\vspace{\baselineskip}
\noindent \textbf{Daily retraining.}  To validate our practice of
daily retraining, we conducted a randomized controlled trial of
several ``out-of-date'' versions of the TTP on the \website website
between Aug.~7 and Aug.~30, 2019. We compared versions of the TTP
trained in February, March, April, and May, compared with the ``live''
TTP that is retrained each day.  We collected between 106 and 131
stream-days of data for each of these schemes.  Somewhat to our
surprise, we were not able to detect a significant difference in
performance between any of these ABR schemes---even comparing the
``February'' TTP against one that was retrained each day in August. We
certainly see a benefit from learning \emph{in situ}
(Figure~\ref{fig:emulation_qoe} shows catastrophic behavior from a TTP
learned in a network emulation), but our practice of daily retraining in
situ appears to be overkill.

\section{Experimental Results}
\label{s:evaluation}

We now present findings from our experiments with the \website study,
including the evaluation of \name. Our main results are shown in
Figure~\ref{fig:overall}.
In summary, we conducted a parallel-group, blinded-assignment,
randomized controlled trial of five ABR schemes between Jan.~19 and
Aug.~7, and between Aug.~30 and Sept.~12, 2019 (the cutoff for
this submission). The data include 8.5 stream-years of data split
across five algorithms, counting all streams that played at least 4
seconds of video. A standardized diagram of the experimental flow is
available in the appendix (Figure~\ref{fig:consort}).

We found that old-fashioned ``buffer-based'' control performs
surprisingly well, despite its status as a frequently outperformed
research baseline. In the overall dataset (Figure~\ref{fig:overall},
left-hand side), BBA outperforms MPC-HM on stalls and is statistically
indistinguishable on video quality. It outperforms Pensieve on quality
and is indistinguishable on stalls. RobustMPC-HM has an even lower
stall rate, at the cost of considerable loss of quality. Among
``slow'' network paths (those with average throughput less than
6~Mbit/s; see Figure~\ref{fig:overall}, right-hand side), Pensieve
shows an advantage in reducing the stall rate, again at a considerable
cost in quality.

\begin{figure}[t]
\centering
\includegraphics[width=0.92\columnwidth]{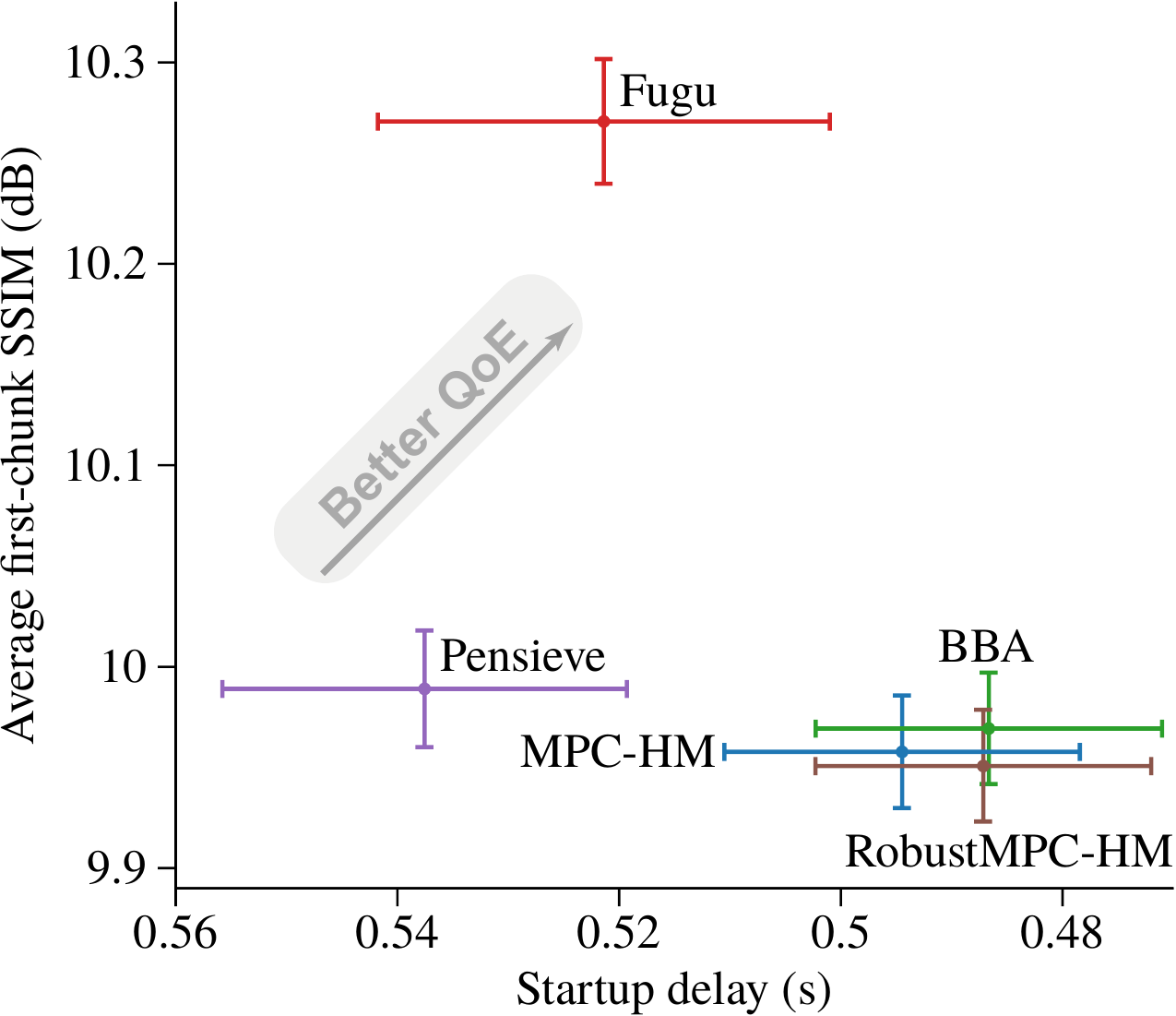}
\caption{\rm On a cold start, \name's ability to bootstrap ABR
  decisions from congestion-control statistics (e.g., RTT) boosts
  initial quality.}
\label{fig:first_chunk}
\end{figure}

The only scheme to consistently outperform BBA in both stalls and
quality was \name, but only when \emph{all} features of the TTP were
used. If we remove the probabilistic ``fuzzy'' nature of \name's
predictions, \emph{or} the ``depth'' of the neural network, \emph{or}
the prediction of transmission time as a function of chunk size (and
not simply throughput), \name forfeits its advantage
(\S\S\ref{ss:ablation}). \name also outperformed other schemes in
terms of SSIM variability (Figure~\ref{fig:summary}). The TTP's
use of low-level TCP statistics was helpful
on a cold start to a new session; this allowed \name to begin at a
higher quality (Figure~\ref{fig:first_chunk}).

We conclude that robustly beating ``simple'' algorithms with machine
learning may be surprisingly difficult, notwithstanding promising
results in contained environments such as simulators and
emulators. The gains that learned algorithms have in optimization or
smarter decision making may come at a tradeoff in brittleness or
sensitivity to heavy-tailed behavior.

\subsection{\name users streamed for longer}

\begin{figure}[t]
\begin{centering}
\includegraphics[width=0.98\columnwidth]{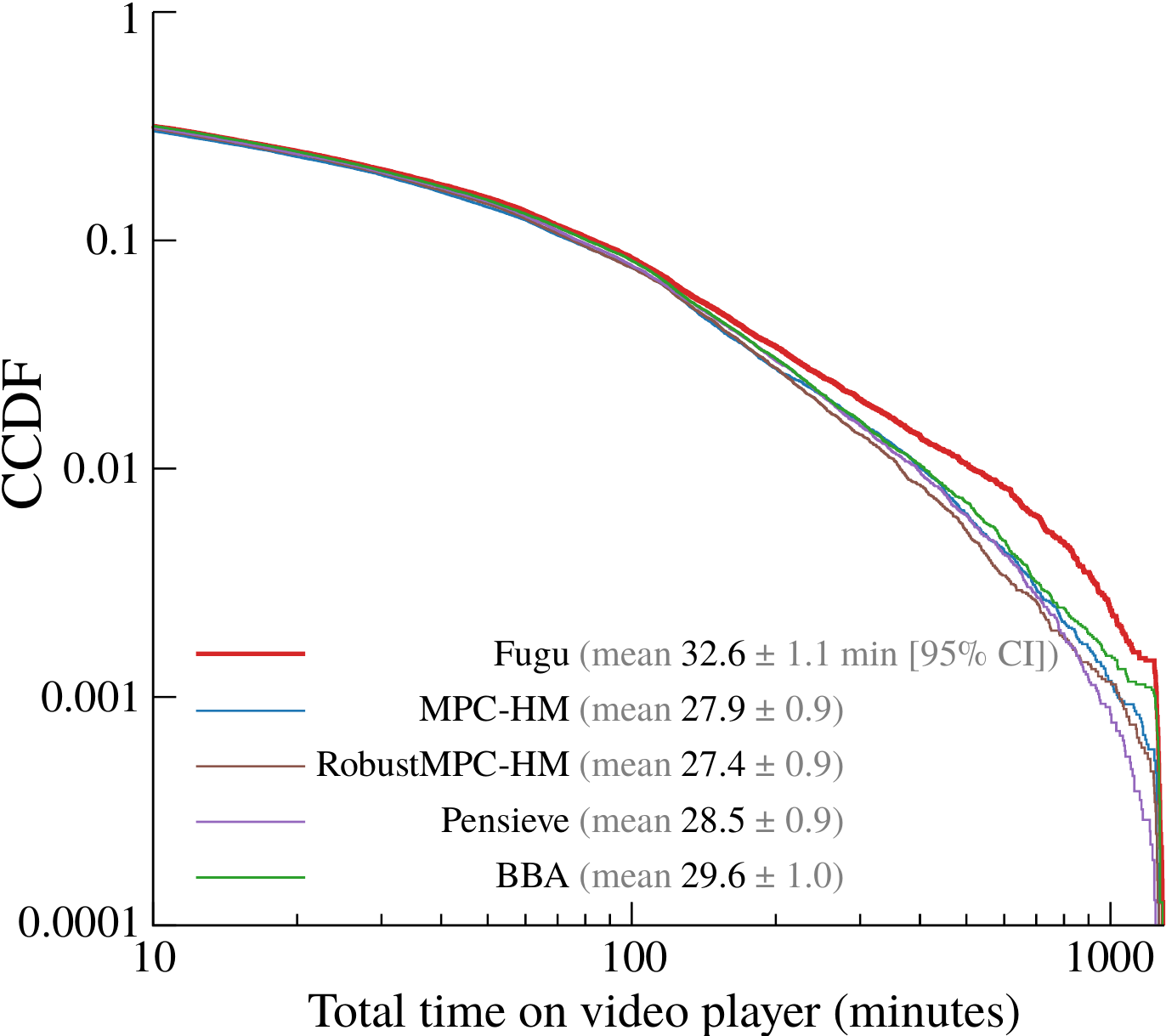}

\end{centering}

\caption{\rm Users randomly assigned to \name chose to remain
  on the \website video player about 10\%--20\% longer, on average,
  than those assigned to other schemes. Users were blinded to the
  assignment. This average difference was driven solely by the upper 5\% tail
  (sessions lasting more than 2.5~hours). Time-on-site is a
  figure of merit in the industry and may correlate with QoE, but we
  do not fully understand this effect.}
\label{fig:watchtime}
\label{tab:session_durations}
\end{figure}

  We observed significant differences in the session durations of
  users across algorithms (Figure~\ref{tab:session_durations}). Users
  whose sessions were assigned to \name chose to remain on the
  \website video player about 10--20\% longer, on average, than those
  assigned to other schemes.  Users were blinded to the assignment,
  and we believe the experiment was carefully executed not to ``leak''
  details of the underlying scheme (MPC and \name even share most of
  their codebase). The average difference was driven solely by the
  upper 5\% tail of viewership duration (sessions lasting more than
  2.5 hours)---viewers assigned to \name are much more likely to keep
  streaming beyond this point, even as the distributions are nearly
  identical until then.

  Time-on-site is a figure of merit in the video-streaming industry
  and might be increased by delivering better-quality video, but we
  simply do not know enough about what is driving this phenomenon.

\subsection{The benefits of learning \emph{in situ}}

\begin{figure*}
\begin{subfigure}{0.62\columnwidth}
  \centering
  \includegraphics[height=0.85\textwidth]{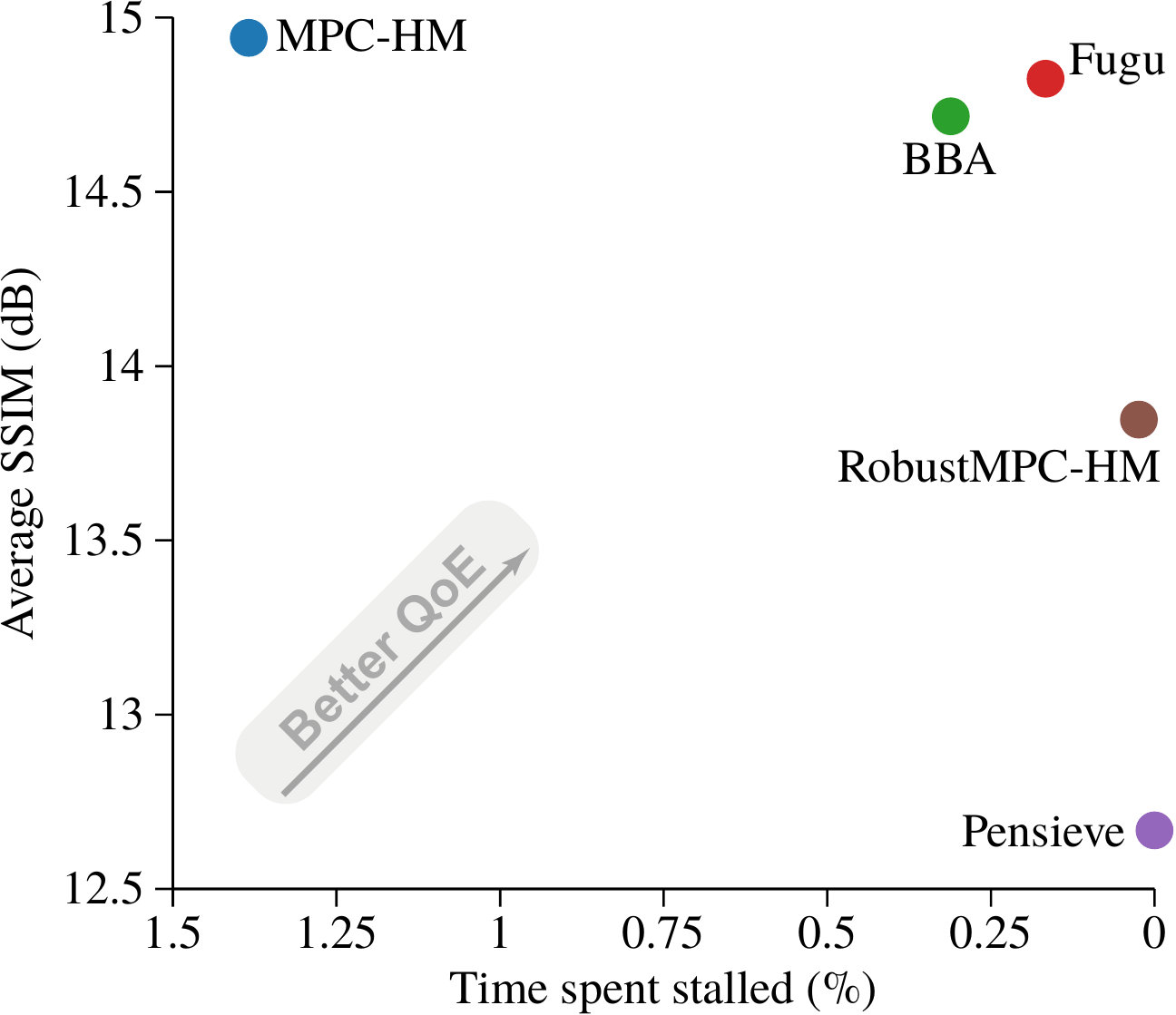}
\end{subfigure}
\hspace{0.4cm}
\begin{subfigure}{0.62\columnwidth}
  \centering
  \includegraphics[height=0.85\textwidth]{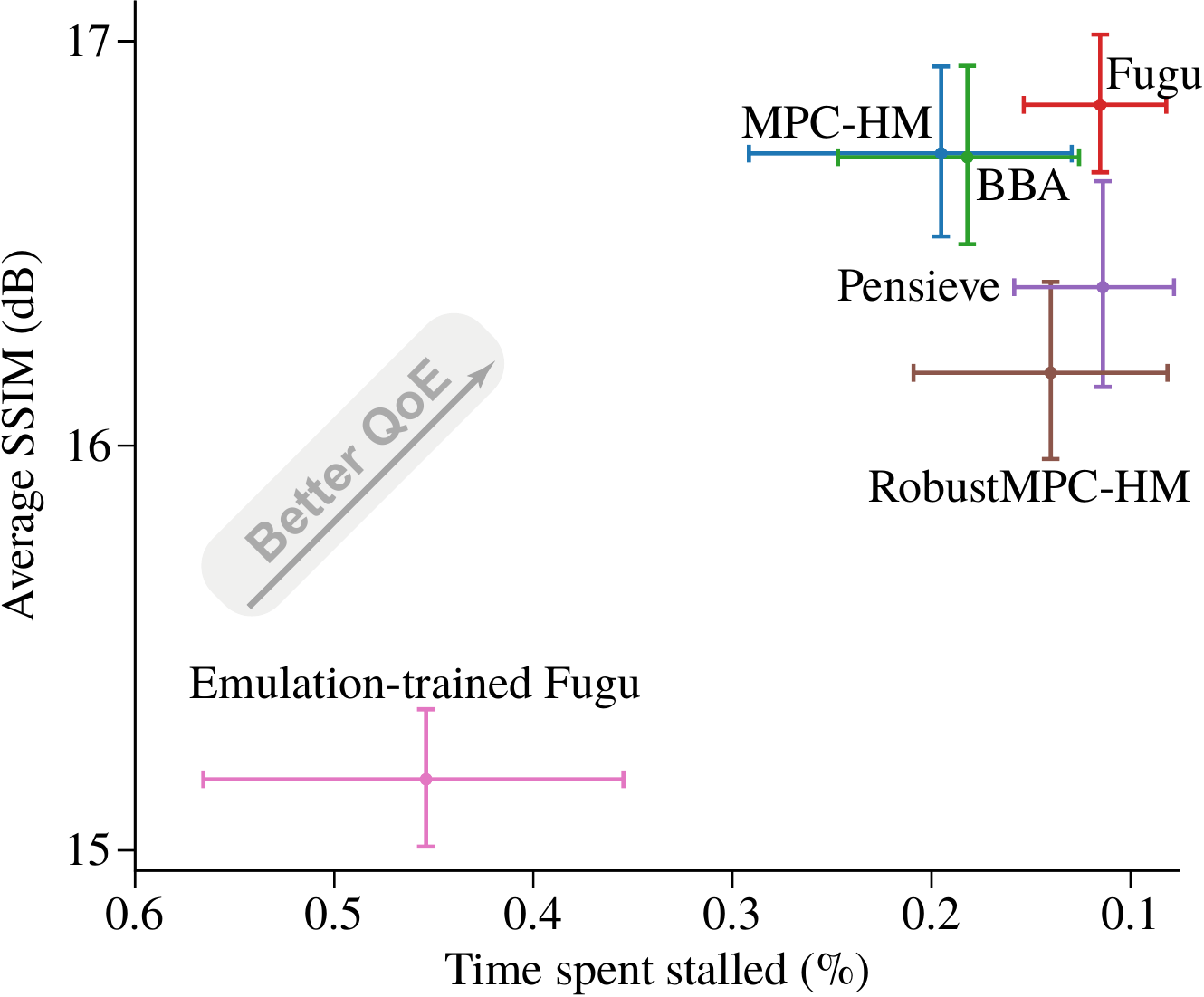}
\end{subfigure}
\hspace{0.4cm}
\begin{subfigure}{0.62\columnwidth}
  \centering
  \includegraphics[height=0.85\textwidth]{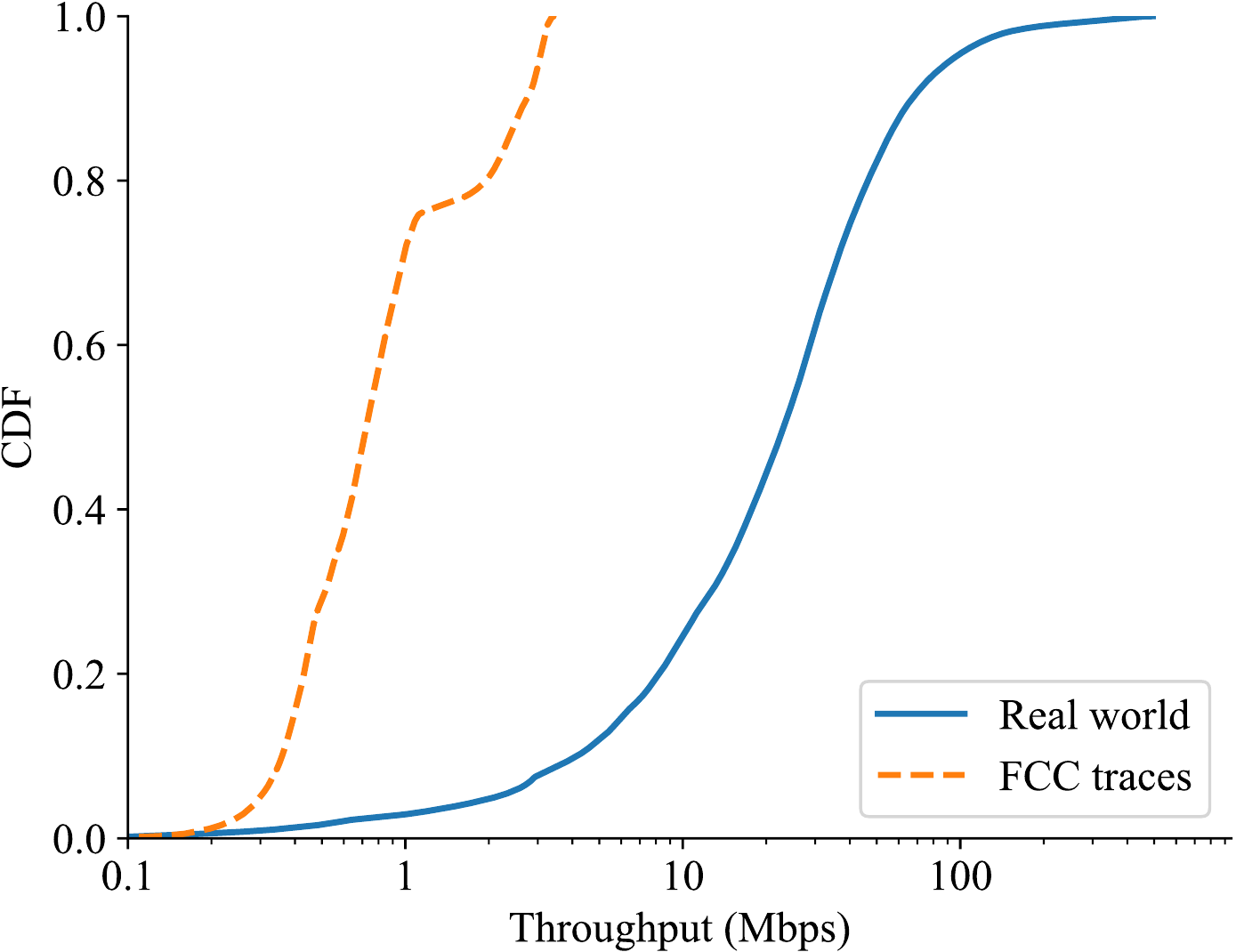}
\end{subfigure}
\caption{\rm \textbf{Left:} performance in emulation, run in
  mahimahi~\cite{mahimahi} using the FCC traces~\cite{fccdata},
  following the method of Pensieve~\cite{pensieve}. \textbf{Middle:}
  experimental results from \website. During Jan.--April 2019, we randomized
  sessions to a set of algorithms
  including ``emulation-trained \name,'' and
  show results of 45,695 streams, 328 stream-days of data from this time period.
  Training on these traces did not generalize to the real-world setting.
  \textbf{Right:} comparison of throughput distribution on FCC traces and
  \website, estimated using the experimental results from the left two figures.}
\label{fig:emulation_qoe}
\end{figure*}

Each of the ABR algorithms we evaluate was
evaluated in emulation in prior works~\cite{pensieve, mpc}. Notably,
the results in those works are qualitatively different from some of
the real world results we have seen here---for example, buffer-based
control outperforming MPC-HM and Pensieve.

  To investigate this further, we constructed an emulation environment similar
  to that used in~\cite{pensieve}. This involved running the \website media
  server locally, and launching headless Chrome clients inside
  mahimahi~\cite{mahimahi} shells to connect to the server. Each mahimahi shell
  imposed a 40~ms end-to-end delay on traffic originating inside it and limited
  the downlink capacity over time to match the capacity recorded in a set of FCC
  broadband network traces~\cite{fccdata}. As in the Pensieve evaluation, uplink speeds
  in all shells were capped at 12 Mbps. Within this test setup, we automated 12
  clients to repeatedly connect to the media server, which would play a 10
  minute clip recorded on NBC over each network trace in the dataset. Each
  client was assigned to a different combination of ABR and CC algorithms, and
  played the 10 minute video repeatedly over more than 15 hours of FCC traces.
  The results from this experiment are depicted in
  Figure~\ref{fig:emulation_qoe}.

  We trained a version of \name in this emulation environment to
  compare its performance with a version trained \emph{in situ} (on
  data from \website). Compared with the \emph{in situ} \name---or
  with every other ABR scheme---the real-world performance of
  emulation-trained \name was horrible
  (Figure~\ref{fig:emulation_qoe}, middle panel).
  Looking at the other ABR schemes, it is illuminating to comparing
  Figure~\ref{fig:emulation_qoe} with Figure~\ref{fig:overall}---the
  emulation results differ markedly from the real world.  In emulation
  (left side of figure), almost every algorithm tested lies somewhere
  along the SSIM/stall frontier, with Pensieve rebuffering the least
  and MPC delivering the highest quality video, and the other
  algorithms lying somewhere in between. In the real experiment
  (middle of figure), we see a more muddled picture, with \name
  apparently outperforming other algorithms in both quality and stall
  time. These results suggest research opportunities in constructing
  network emulators that capture additional dynamics of the real
  Internet.
  

\subsection{Remarks on Pensieve and RL for ABR} \label{ss:pensieve-performance}



  The original Pensieve
  paper~\cite{pensieve} demonstrated that Pensieve outperformed MPC-HM, RobustMPC-HM, and BBA
  in both simulation-based tests and in video streaming
  tests on a low and high speed real-world networks. Our results differ;
  we believe the mismatch may have occurred for several reasons.

  First, we have found that emulation-based training and testing (or,
  at least, mahimahi tests with the FCC dataset) do not capture the
  vagaries of the real-world paths seen in the \website study. Unlike
  real-world randomized trials, trace-based emulators and
  simulators allow experimenters to run two different algorithms on
  the \emph{same} conditions, eliminating the effect of the play of
  chance in giving different algorithms a different distribution of
  watch times, network behaviors, etc. However, it is difficult to
  characterize the \emph{systematic} uncertainty that comes from
  selecting a set of traces that may omit the variability or
  heavy-tailed nature of a real deployment experience (both network behaviors
  as well as user behaviors, such as watch duration).

  Reinforcement learning (RL) schemes such as Pensieve may be at a
  particular disadvantage from this phenomenon. Unlike supervised
  learning schemes that can learn from training ``data,''
  reinforcement learning requires a training \emph{environment} to
  respond to a sequence of control decisions and decide on the
  appropriate consequences and reward.  That environment could be real
  life instead of a simulator, but the level of statistical noise we
  observe would make this type of learning extremely slow or require
  an extremely broad deployment of algorithms in training. Reinforcement learning relies on
  being able to slightly vary a control action and detect a change in the resulting reward. By our calculations, the
  variability of inputs is such that it takes about 2 stream-years of
  data to reliably distinguish two ABR schemes whose innate ``true''
  performance differs by 15\%. To make RL practical, future work may
  need to explore techniques to reduce this
  variability~\cite{reducingvariance} or construct more faithful
  simulators and emulators that model tail behaviors~\cite{pantheon}.

  Second, most of the evaluation of Pensieve in the original paper
  focused on training and evaluating Pensieve using a single test
  video. As a result, the state space that model had to explore was
  inherently more limited.  Evaluation of the Pensieve ``multi-video
  model''---which we have to use for our experimental setting---was more
  limited.  Our results are more consistent with a recent large-scale
  study of a Pensieve-multi-video-like scheme on 30 million streams at
  Facebook~\cite{realworldpensieve}; this study found a small benefit
  in bitrate over Facebook's default ABR scheme, and no significant
  benefit in rebuffering.

  Finally, Pensieve optimizes a QoE metric centered around bitrate as
  a proxy for video quality. Altering this would have required
  significant surgery to provide new values to the Pensieve neural
  network over time. Figure~\ref{fig:bitrate} shows that Pensieve was
  the \#2 scheme in terms of bitrate (below BBA) in the primary
  analysis.
  We emphasize that our findings do not indicate that Pensieve
  cannot be a useful ABR algorithm, especially in a scenario where similar, pre-recorded
  video is played over a familiar set of known networks.


\section{Limitations}
\label{s:limits}

The design of the \website experiment and the \name system are subject
to important limitations that may affect their performance and
generalizability.

\subsection{Limitations of the experiments}

Our randomized controlled trial represents a rigorous, but necessarily
``black box,'' study of ABR algorithms for video streaming. We don't
know the true distribution of network paths and throughput-generating
processes; we don't know the participants or why the distribution in watch times differs
by assigned algorithm; and we don't know
how to emulate these behaviors accurately in a controlled environment.

We have supplemented this black-box work with ablation
analyses to relate the real-world performance of \name to the $l^2$
accuracy of its predictor, and have studied various ablated versions of
\name in deployment. However, ultimately part of the reason for this
paper is that we \emph{cannot} replicate the experimental findings
outside the real world---a real world whose behavior is noisy and
takes lots of time to measure precisely. That may be an unsatisfying
conclusion, and we doubt it will be the final word on this
topic. Perhaps it will become possible to model enough of the vagaries of the
real Internet ``in silico'' to enable the development of robust
control strategies without extensive real-world experiments.

It is also unknown to what degree \website's results---which are about
a single server with 10~Gbps connectivity in a well-provisioned
datacenter, sending to clients across our entire country over the
wide-area Internet---generalize to a different server at a different
institution, much less the more typical paths between a user on an
access network and their nearest CDN edge node. We don't know for sure
if the pre-trained \name model would work in a different location, or
whether training a new \name based on data from that server would
yield comparable results. Our results show that learning \emph{in
  situ} works, but we don't know how specific the \emph{situs} needs
to be.

Although we believe that past research papers may have underestimated
the uncertainties in real-world measurements with realistic Internet
paths and users, we also may be guilty of underestimating our own
uncertainties or---also possible---emphasizing uncertainties that are
only relevant to small or medium-sized academic studies, such as ours,
and irrelevant to the industry. The current load on \website is about
50 concurrent streams on average, meaning we collect about 50
stream-days of data per day. Our primary analysis covers about 1.7
stream-years of data per scheme collected over a seven-month period,
and was sufficient to measure its performance metrics to within about
$\pm 15\%$ (95\% CI).

By contrast, we understand YouTube has an average load over 50
\emph{million} concurrent streams at any given time. We imagine the
considerations of conducting data-driven experiments at this level may
be completely different---perhaps less about statistical uncertainty,
and more about systematic uncertainties and the difficulties of
running experiments and accumulating so much data. (We also understand
that YouTube is only able to measure its own performance once every 48
hours because of the vastness of data that needs to be aggregated.)


Some of \name's performance (and that of MPC, RobustMPC, and BBA)
relative to Pensieve may be due to the fact that these four schemes
received more information as they ran---namely, the SSIM of each
possible version of each future chunk---than did Pensieve. It is
possible that an ``SSIM-aware'' Pensieve might perform better. We
tried to approximate this sort of scheme (an SSIM-aware neural network
trained in emulation) with the ``Emulation-trained \name'' benchmark
(Figure~\ref{fig:emulation_qoe}).  The load of calculating SSIM for
each encoded chunk is not insignificant---about an extra 40\% on top
of the cost of encoding the video.

\subsection{Limitations of \name}

There is a sense that data-driven algorithms that more ``heavily''
squeeze out performance gains may also put themselves at risk to
brittleness when a deployment environment drifts from one where the
algorithm was trained. In that sense, it is hard to say whether
\name's performance might decay catastrophically some day. We tried
and failed to demonstrate a quantitative benefit from \emph{daily}
retraining over ``every six months'' retraining, but at the same time,
we cannot be sure that some surprising detail tomorrow---e.g.,~a new
user from an unfamiliar network---won't send \name into a tailspin
before it can be retrained. Our eight months of data on a growing
userbase suggests, but doesn't guarantee, robustness to a changing
environment.

\name does not consider several issues that other research has
concerned itself with---e.g., being able to ``replace''
already-downloaded chunks in the buffer with higher quality
versions~\cite{sabre}, or optimizing the joint QoE of multiple clients
who share a congestion bottleneck.

\name is not tied as tightly to the TCP or congestion control as it
might be---for example, \name could wait to send a chunk until the TCP
sender tells it that there is a sufficient congestion window for most of
the chunk (or the whole chunk) to be sent immediately. Otherwise, it
\emph{might} choose to wait and make a better-informed decision
later. \name does not schedule the transmission of chunks---it will
always send the next chunk as long as the client has room in its
playback buffer.

\section{Conclusion}
\label{s:conclusion}

Machine-learned systems in computer networking sometimes describe
themselves as achieving near-``optimal'' performance, based on results
in a contained or modeled version of the problem~\cite{learnability,
  bola, pensieve}.

In this paper, we suggest that these efforts can benefit from
considering a broader notion of performance and optimality. Good, or
even near-optimal, performance in a simulator or emulator does not
necessarily predict good performance over the wild Internet, with its
variability and heavy-tailed distributions. It remains a challenging
problem to gather the appropriate training data (or in the case of RL
systems, training environment) to properly learn and validate such systems.

In this paper, we asked: \emph{what does it take to create a learned
  ABR algorithm that robustly performs well over the wild Internet?}
In effect, our best answer is to cheat: train the algorithm \emph{in
  situ} on data from the real deployment environment, and use an
algorithm whose structure is sophisticated enough (a neural network)
and yet also simple enough (a predictor amenable to supervised
learning on data, feeding a classical controller) to benefit from that
kind of training.

Over the last eight months, we have streamed 14.2 years of video to
56,000 users across the Internet. Sessions are randomized in blinded
fashion among algorithms, and client telemetry is recorded for
analysis. The \name algorithm robustly outperformed other schemes, both
simple and sophisticated, on objective measures (SSIM, stall time, SSIM
variability) and increased the duration that users chose to continue
streaming.\footnote{We do not claim this as an accomplishment per se.}

We have found the \website approach an enormously powerful tool for
networking research---it is very fulfilling to be able to ``measure,
then build''~\cite{measurefirst} to iterate rapidly on new ideas and
gain feedback. Accordingly, we are opening \website as an ``open
research'' platform. Along with this paper, we are publishing our full
archive of traces and results on the \website website. The system
posts new data each day, along with a summary of results from the
ongoing experiments, with confidence intervals similar to those in
this paper. (The format is described in the appendix.) We will redact
some fields from the public archive (e.g., IP address and user ID) but
are willing to work with researchers in the community on access to
this data as appropriate. \website and \name are also open-source
software.\footnote{\url{https://github.com/StanfordSNR/puffer}}

We plan to operate \website for several years and invite researchers
to train and validate new algorithms for ABR control, network and throughput
prediction, and congestion control on its traffic. We believe that
\website could serve as a helpful ``medium-scale'' stepping-stone for
new algorithms, partway between the flexibility of network emulation
and the vastness of data---but also conservatism about deploying new
algorithms---of commercial services. We are eager to collaborate with
and learn from the community's ideas on how to design and deploy
robust learned systems for the Internet.

\label{beforerefs}

\bibliographystyle{plain}
\bibliography{reference}

\newpage
\appendix

\renewcommand\thefigure{\thesection\arabic{figure}}
\setcounter{figure}{0}

\begin{sidewaysfigure*}

    \section{Randomized trial flow diagram}
    \centering

    \includegraphics[width=\textheight]{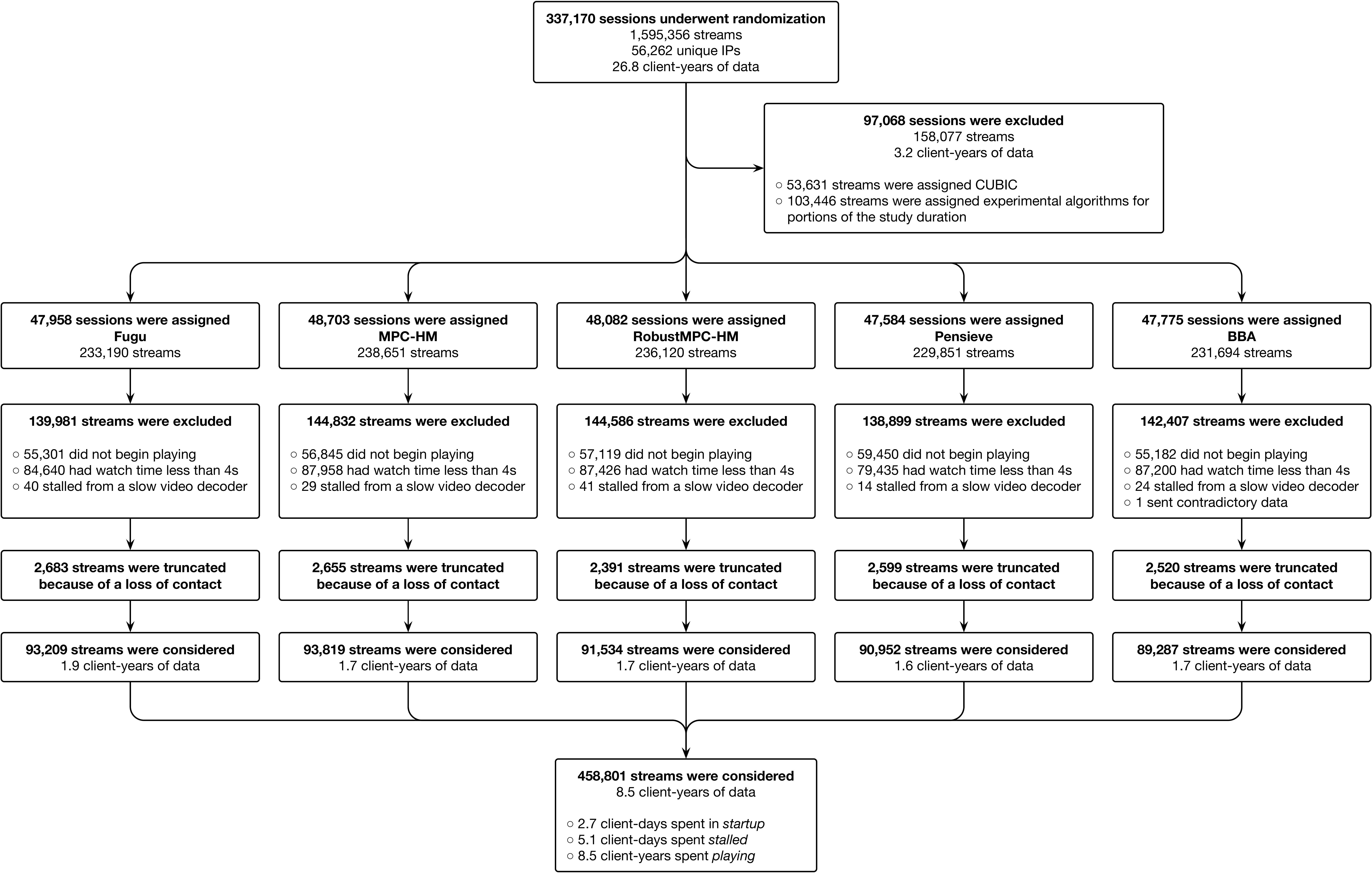}

    \hspace{0.5ex}

    \caption{\rm CONSORT-style diagram~\cite{schulz2010consort} of
    experimental flow for the primary results
    (Figures~\ref{fig:summary} and \ref{fig:overall}), obtained during
    the period Jan.~1--Aug.~7, 2019, and Aug.~30--Sept.~12, 2019. A
    ``session'' represents one visit to the \website video player and
    may contain many ``streams.'' Reloading starts a new session, but
    changing channels only starts a new stream and does not change TCP
    connections or ABR algorithms. A large number of streams never
    began playing; these were often users rapidly changing channels or using
    an incompatible browser.}  \label{fig:consort}

\end{sidewaysfigure*}

\clearpage
\section{Description of open data}


The open data we release comprise different sets of data -- ``measurements'' --
with each measurement containing a piece of information from a video server or
a client. Below we highlight the format of interesting fields in
three measurements essential for analysis of this data:
\verb|video_sent|, \verb|video_acked|, and \verb|client_buffer|.

\verb|video_sent| collects a data point every time an 
\website server sends a video chunk to a client. Each data point contains:
\begin{itemize}
  \item \verb|time|: epoch time when the chunk is sent.
  \item \verb|stream_id|: a unique ID to identify a video stream.
  \item \verb|expt_id|: a unique ID for each experimental group. \verb|expt_id|
    can be used to retrieve the configuration (e.g., ABR, congestion control)
    tested in the corresponding experimental group.
  \item \verb|size|: size of the chunk.
  \item \verb|ssim_index|: SSIM of the chunk.
  \item \verb|cwnd|: current congestion window size (\verb|tcpi_snd_cwnd|) in \verb|tcp_info|.
  \item \verb|in_flight|: number of unacknowledged packets in flight
    (\verb|tcpi_unacked| $-$ \verb|tcpi_sacked| $-$ \verb|tcpi_lost| $+$ \verb|tcpi_retrans|).
  \item \verb|min_rtt|: minimum RTT (\verb|tcpi_min_rtt|).
  \item \verb|rtt|: smoothed RTT estimate (\verb|tcpi_rtt|). 
  \item \verb|delivery_rate|: estimate of TCP throughput (\verb|tcpi_delivery_rate|).
\end{itemize}

\verb|video_acked| collects a data point every time an 
\website server receives a video chunk acknowledgement from a client.
Each data point can be matched to a data point in \verb|video_sent| (if the chunk
is ever acknowledged) and used to calculate the transmission time of the chunk.

\verb|client_buffer| collects client-side buffer information
on a regular interval and when certain events occur. Each data point contains:
\begin{itemize}
  \item \verb|event|: event type, e.g., was this triggered by a regular report every quarter
    second, or because the client stalled or began playing.
  \item \verb|buffer|: playback buffer size.
  \item \verb|cum_rebuf|: cumulative rebuffer time in the current stream.
\end{itemize}

Between January 1 and September 12, 2019, we collected 261,280,238 data points
in \verb|video_sent|, 264,374,831 data points in \verb|video_acked|, and
1,729,618,482 data points in \verb|client_buffer|.

\end{document}